\documentclass[preprint]{elsarticle}
\usepackage{placeins}
\usepackage{amsmath}
\usepackage{amssymb}
\usepackage{tikz}
\usepackage{graphicx}
\usepackage{epstopdf}
\usepackage{natbib}

\begin{document}

\title{Weak dissipation drives and enhances Wada basins in three-dimensional chaotic scattering \tnoteref{t1}}
\tnotetext[t1]{This work has been supported by the Spanish State Research Agency (AEI) and the European Regional Development Fund (ERDF, EU) under Project No.~PID2019-105554GB-I00.}

\author[1]{Diego S. Fern\'{a}ndez}

\author[1]{Jes\'{u}s M. Seoane \corref{cor1}}
\cortext[cor1]{Corresponding author}
\ead{jesus.seoane@urjc.es}

\author[1,2]{Miguel A.F. Sanju\'{a}n}

\address[1]{Nonlinear Dynamics, Chaos and Complex Systems Group, Departamento de F\'{i}sica, Universidad Rey Juan Carlos \\ Tulip\'{a}n s/n, 28933 M\'{o}stoles, Madrid, Spain}

\address[2]{Department of Applied Informatics, Kaunas University of Technology \\ Studentu 50-415, Kaunas LT-51368, Lithuania}

\begin{abstract}
Chaotic scattering in three dimensions has not received as much attention as in two dimensions so far. In this paper, we deal with a three-dimensional open Hamiltonian system whose Wada basin boundaries become non Wada when the critical energy value is surpassed in the absence of dissipation. In particular, we study here the dissipation effects on this topological change, which has no analogy in two dimensions. Hence, we find that non-Wada basins, expected in the absence of dissipation, transform themselves into partially Wada basins when a weak dissipation reduces the system energy below the critical energy. We provide numerical evidence of the emergence of the Wada points on the basin boundaries under weak dissipation. According to the paper findings, Wada basins are typically driven, enhanced and, consequently, structurally stable under weak dissipation in three-dimensional open Hamiltonian systems.
\end{abstract}


 \maketitle

\section{Introduction} \label{sec:1}

Chaotic scattering \cite{seoane2013} is an important field in nonlinear dynamics with a wide range of different applications in physics, such as fluid dynamics \cite{daitche2014} or material transitions \cite{toledomarin2018}, but also in biology \cite{scheuring2003} or medicine \cite{schelin2010}, to name a few. A chaotic scattering process is a direct manifestation of transient chaos \cite{tel2015} in open dynamical systems. In these processes, which can be modeled by means of open Hamiltonian systems, typically a point particle interacts with a nonlinear potential function for a finite time, or transient, until it escapes through one of the possible exits to infinity. The exit basins, sets of initial conditions that lead to a certain exit, are a common tool in this kind of systems to ascertain which initial condition regions are unpredictable, since a minimum uncertainty in the initial conditions can hinder the exit prediction. The vast majority of works on open Hamiltonian systems have been devoted to conservative two-dimensional problems \cite{zotos2015, navarro2021}. Nonetheless, these systems have also been studied in the presence of external perturbations, such as periodic driving \cite{nieto2018}, additive noise \cite{nieto2021} or weak dissipation \cite{motter2001, seoane2006, seoane2007}. 

Specifically, in the presence of dissipation, the unpredictability in determining the system final state has been studied, both in discrete maps \cite{motter2001} and two-dimensional open Hamiltonian systems \cite{seoane2006, seoane2007}. These works conclude that the nonhyperbolic dynamics, dominated by Kolmogorov-Arnold-Moser (KAM) tori, is suddenly destroyed by introducing an even extremely small amount of dissipation. As a consequence of this, the more intense the dissipation rate, the less the system unpredictability. On the other hand, the Wada property has been also investigated in two-dimensional open Hamiltonian systems, whose exit basins typically exhibit this property \cite{aguirre2001}. Furthermore, Wada basins have been demonstrated to be structurally stable in the presence of weak dissipation in these systems \cite{seoane2006}. This is an intriguing topological property of some sets, according to which the boundary separating three or more sets is common to all of them \cite{yoneyama1917}. In this manner, on a Wada boundary is always possible to find all other basins on the boundary between two of them, which represents a clear source of system unpredictability. For further details, see Refs.~\cite{aguirre2009, wagemakers2021}, where the state of the art of the Wada property and the current methods to detect it are reviewed.

However, three-dimensional open Hamiltonian systems under dissipation still require to be investigated, since the conservative scenario is the one that has been the focus of attention so far \cite{kovacs2001, lin2013, drotos2014, naik2019}. In this paper, we deal with a three-dimensional open Hamiltonian system with four possible exits \cite{lai2000}, which has a two-dimensional version \cite{demoura2002}. In particular, in the absence of dissipation, a striking basin topological metamorphosis takes place with no analogy in two dimensions: Wada basins become non-Wada basins when the system energy exceeds a certain critical value $E_c$. In this context, we wonder about the topological consequences of introducing dissipation, since the energy loss can determine whether the basins are Wada or not.

The organization of this paper is as follows. In Sec.~\ref{sec:2}, we describe the Hamiltonian dynamics of the model in the conservative and the dissipative cases. In Sec.~\ref{sec:3}, we compute the system unpredictability versus the dissipation parameter by means of the fractal dimension of the chaotic saddle. This fractal dimension is key to understand the Wada basins in the dissipative system. In Sec.~\ref{sec:4}, we propose the dissipative mechanism by which non-Wada basins in the conservative case become Wada. Furthermore, we compute the fraction of Wada points on the basin boundaries as the dissipation parameter is varied. Finally, in the conclusions, we summarize the main findings and relate them to possible applications where they may be of interest.

\section{Model description} \label{sec:2}

\begin{figure*}[h]
	\centering
	
	\includegraphics[width=0.48\textwidth]{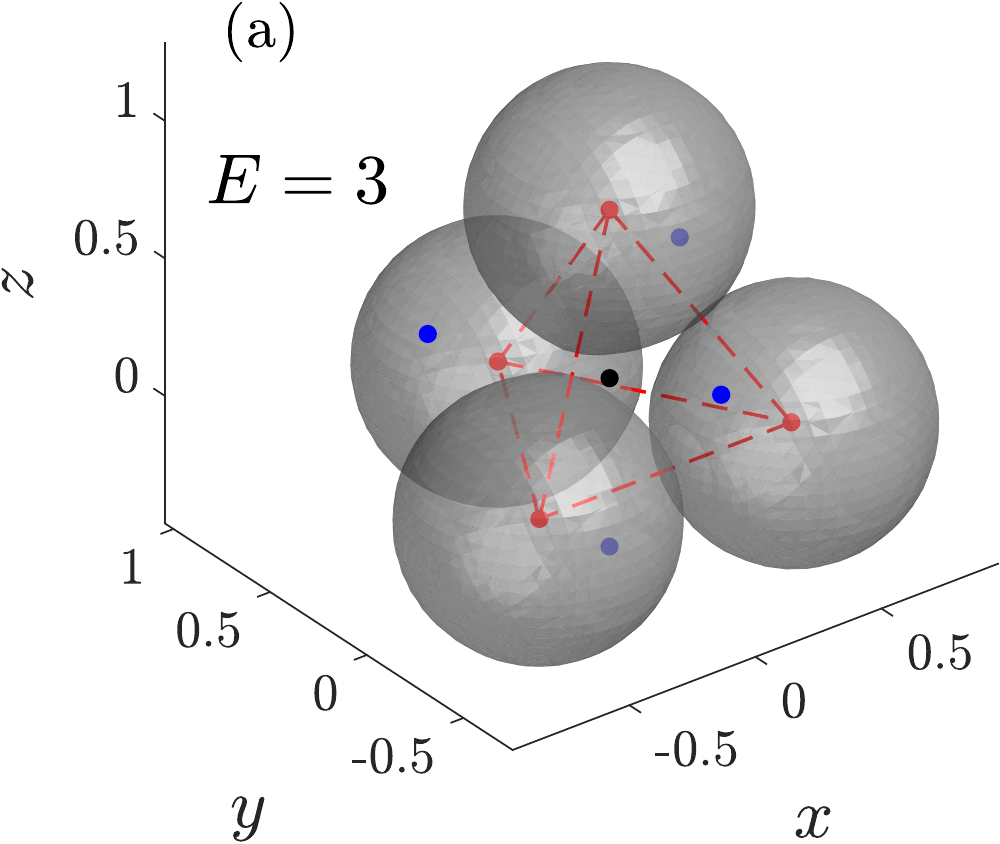}
	\quad
	\includegraphics[width=0.48\textwidth]{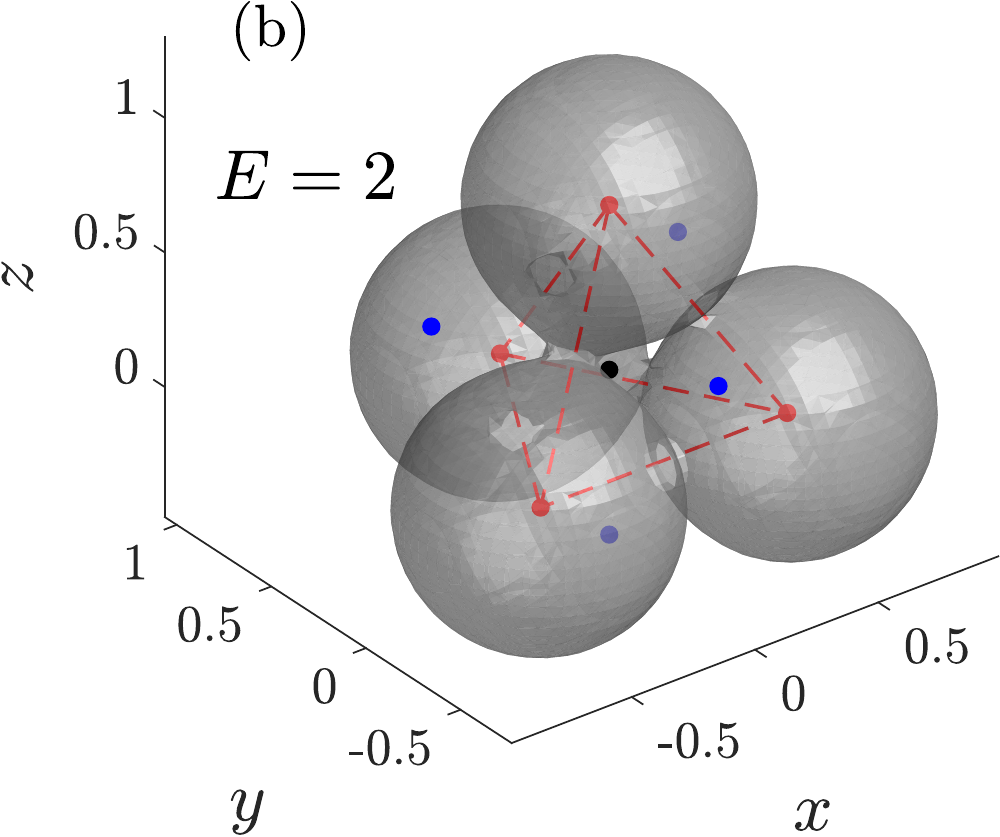}\\
	\bigskip
	\includegraphics[width=0.48\textwidth]{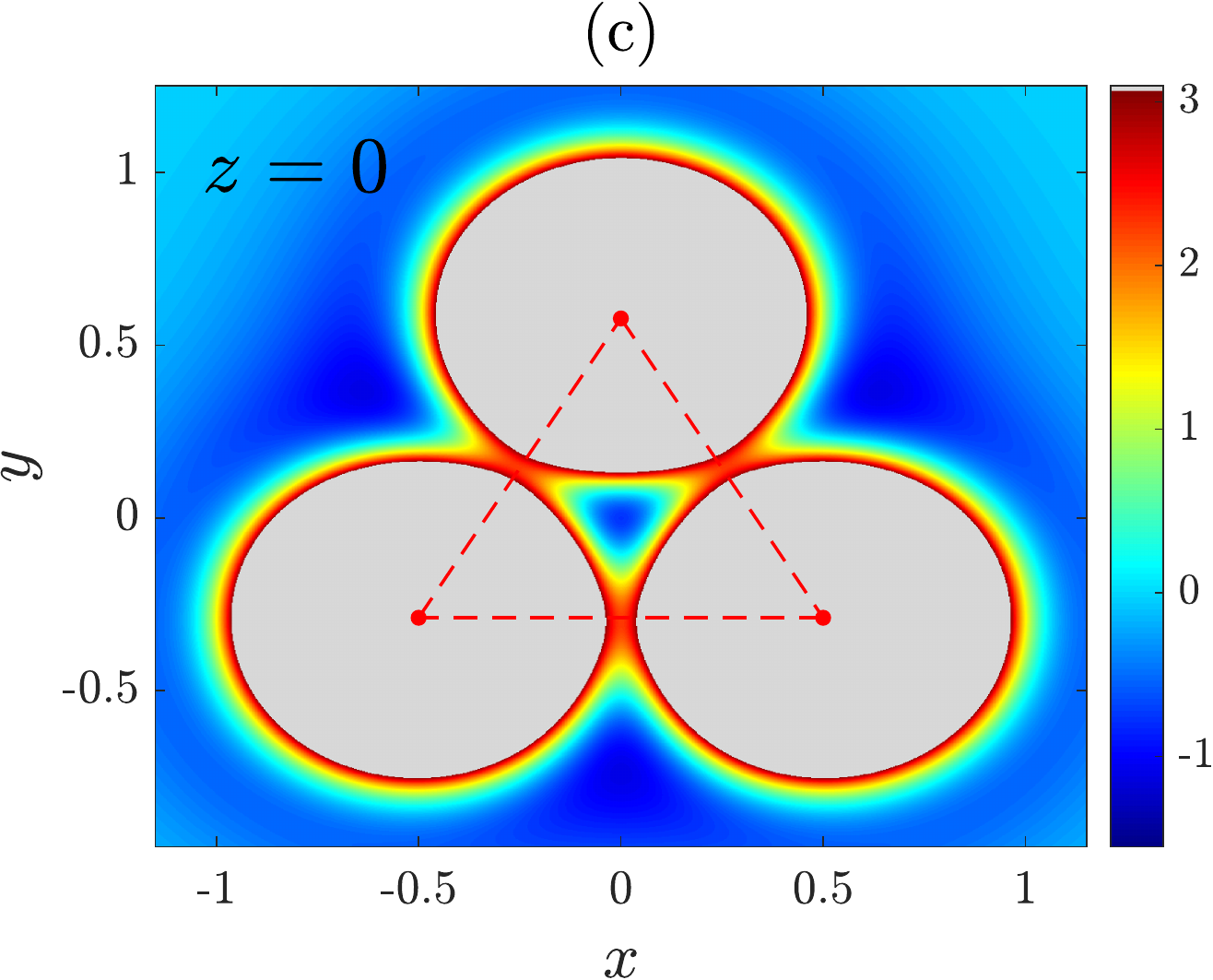}
	\quad
	\includegraphics[width=0.48\textwidth]{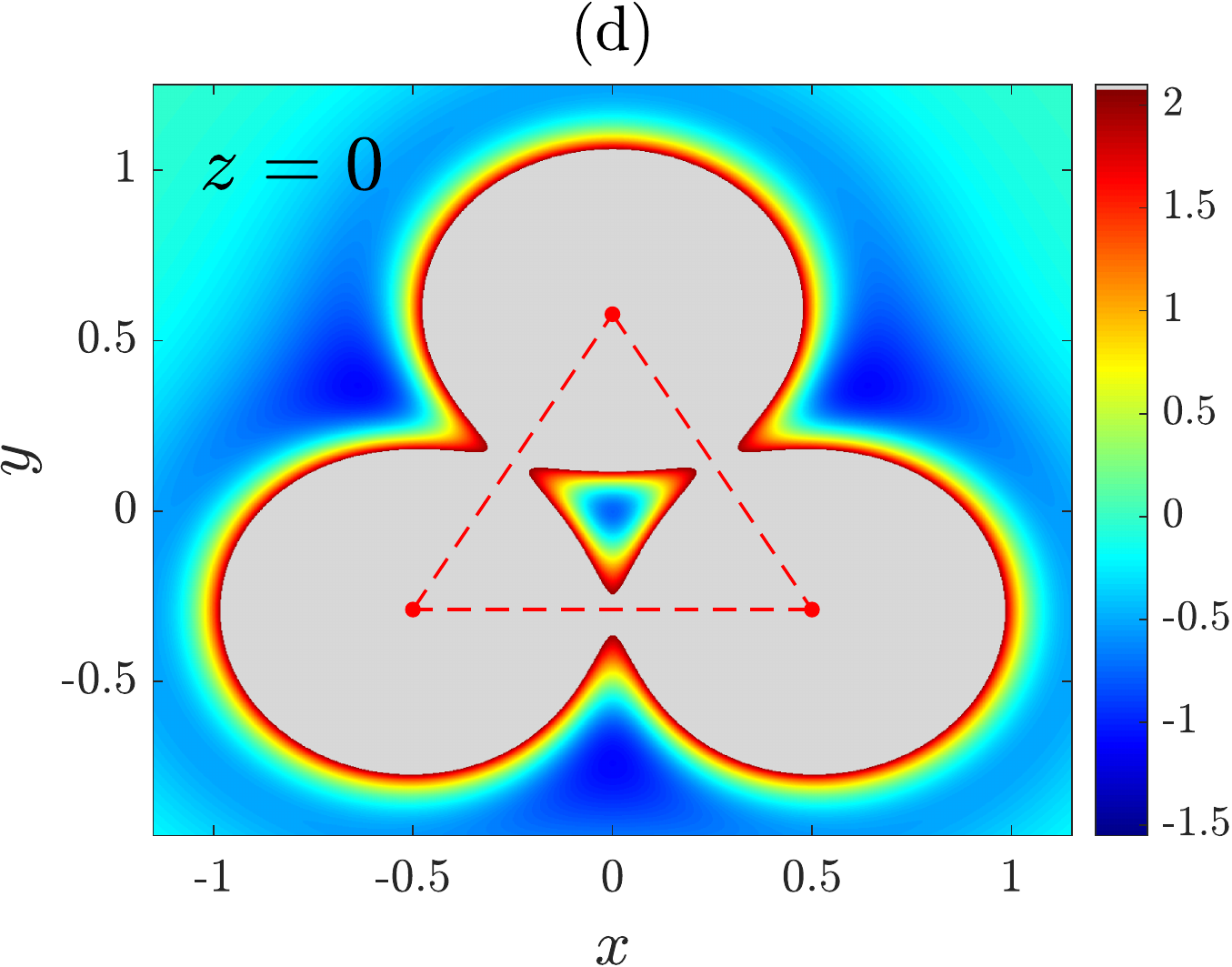}
	
	\caption{The physical space for (a) $E = 3 > E_c$ and (b) $E = 2 < E_c$, where $E_c \approx 2.25$. The interior of the gray spherical-like regions represents the physical space part inaccessible to the particle, which remain (a) disconnected for $E > E_c$ and (b) connected for $E < E_c$. We define the scattering region as the regular tetrahedron (dashed red lines) formed by the four Morse hills’ centers (red dots), since the chaotic behavior is only relevant near the tetrahedron centroid (black dot). Five attractors (black and blue dots) appear when the dissipation is introduced, and the particle may trap in them forever. Finally, we show a potential function section with the plane $z = 0$ for (c) $E = 3$ and (d) $E = 2$, where the minimum value is the same in both cases.}
	\label{fig:1}
\end{figure*}

In order to study the dissipation effects on a three-dimensional chaotic scattering problem, we use a system that models the interaction of a point particle with non-rotating diatomic molecules \cite{lai2000}. The system dynamics is modeled by the dimensionless and time-independent Hamiltonian \begin{equation} H(p_x,p_y,p_z,x,y,z) = \frac{p_x^2 + p_y^2 + p_z^2}{2} + V(x,y,z), \label{eq:1} \end{equation} where $p_x$, $p_y$ and $p_z$ are the momenta and $V(x,y,z)$ is a nonlinear potential function which depends on the positions $x$, $y$ and $z$. The potential consists of four Morse hills \cite{haar1946} whose centers are placed at the vertices of a regular tetrahedron of unit side lengths. Specifically, located at $(x_1, y_1, z_1) = (0, 0, \sqrt{2/3})$, $(x_2, y_2, z_2) = (1/2, -1/(2\sqrt{3}), 0)$, $(x_3, y_3, z_3) = (-1/2, -1/(2\sqrt{3}), 0)$ and $(x_4, y_4, z_4) = (0, 1/\sqrt{3}, 0)$, depicted as red dots in Fig.~\ref{fig:1}. For clarity, the tetrahedron centroid, $(x_c, y_c, z_c) = (0,0,1/\sqrt{24})$, is also depicted as a black dot in Fig.~\ref{fig:1}. Here the three-dimensional Morse hills have spherical symmetry but slightly distorted by the presence of the others. Each one of them consists of a repulsive core (any tetrahedron vertex) surrounded by an attractive region, which is less effective, and ultimately negligible, if the particle moves away from the hill center. The Morse potential represents short-range interactions, implying that the system chaotic behavior is concentrated near the tetrahedron formed by the four hills' centers. In this manner, we define the scattering region as the space bounded by the four tetrahedron planes. That is, we consider that a particle escapes when it crosses any of these planes. The complete potential function is written as \begin{equation} V(x,y,z) = \sum_{i = 1}^4 V_i (x,y,z), \label{eq:2} \end{equation} where \begin{equation} V_i(x,y,z) = \frac{V_0}{2} \left[ 1-e^{-\alpha(r_i - r_e)} \right]^2 - \frac{V_0}{2} \label{eq:3} \end{equation} represents the potential function associated with each one of the four hills, and \begin{equation} r_i(x,y,z) = \sqrt{(x-x_i)^2 + (y-y_i)^2 + (z-z_i)^2} \label{eq:4} \end{equation} represents the distances between the particle position and each one of the four hills' centers. In addition, the quantities $V_0$, $\alpha$ and $r_e$ are the Morse potential parameters, which are related to the well depth, the well steepness and the effective range of each hill, respectively. Without loss of generality, we set along this paper the parameter values as $V_0 = 1$, $\alpha = 6$ and $r_e = 0.68$.

Given a system energy value $E$, there exists a three-dimensional region of the physical space inaccessible to the particle, where $E < V$ is satisfied. That is, the inaccessible region limits are simply the potential barriers where the particle usually bounces off before escaping. In this specific system, when the energy is greater than a certain critical energy $E_c$, the inaccessible region is formed by four disconnected spherical-like regions surrounding the Morse hills' centers, as shown in Fig.~\ref{fig:1}(a). As the system energy is decreased, the four disconnected regions grow in radius, touching each other when the critical energy is reached. Finally, when the energy is smaller than the critical energy, the inaccessible region is a single fully connected region, as observed in Fig.~\ref{fig:1}(b). In addition, we display the potential function in the plane $z = 0$ in Figs.~\ref{fig:1}(c) and \ref{fig:1}(d) to better visualize how inaccessible regions remain connected or disconnected depending on the energy. The existence of this critical energy is key to the phase space topology, even in presence of dissipation, as we explain in detail below. We compute $E_c \approx 2.250414$ by using the critical trajectory concept, which has been developed in two-dimensional open Hamiltonian systems \cite{fernandez2021}. Hence, we work along the paper on a range of initial energy values $E_0 \in [2,3]$, which includes the critical energy.

\begin{figure*}[h]
	\centering
	
	\includegraphics[width=0.48\textwidth]{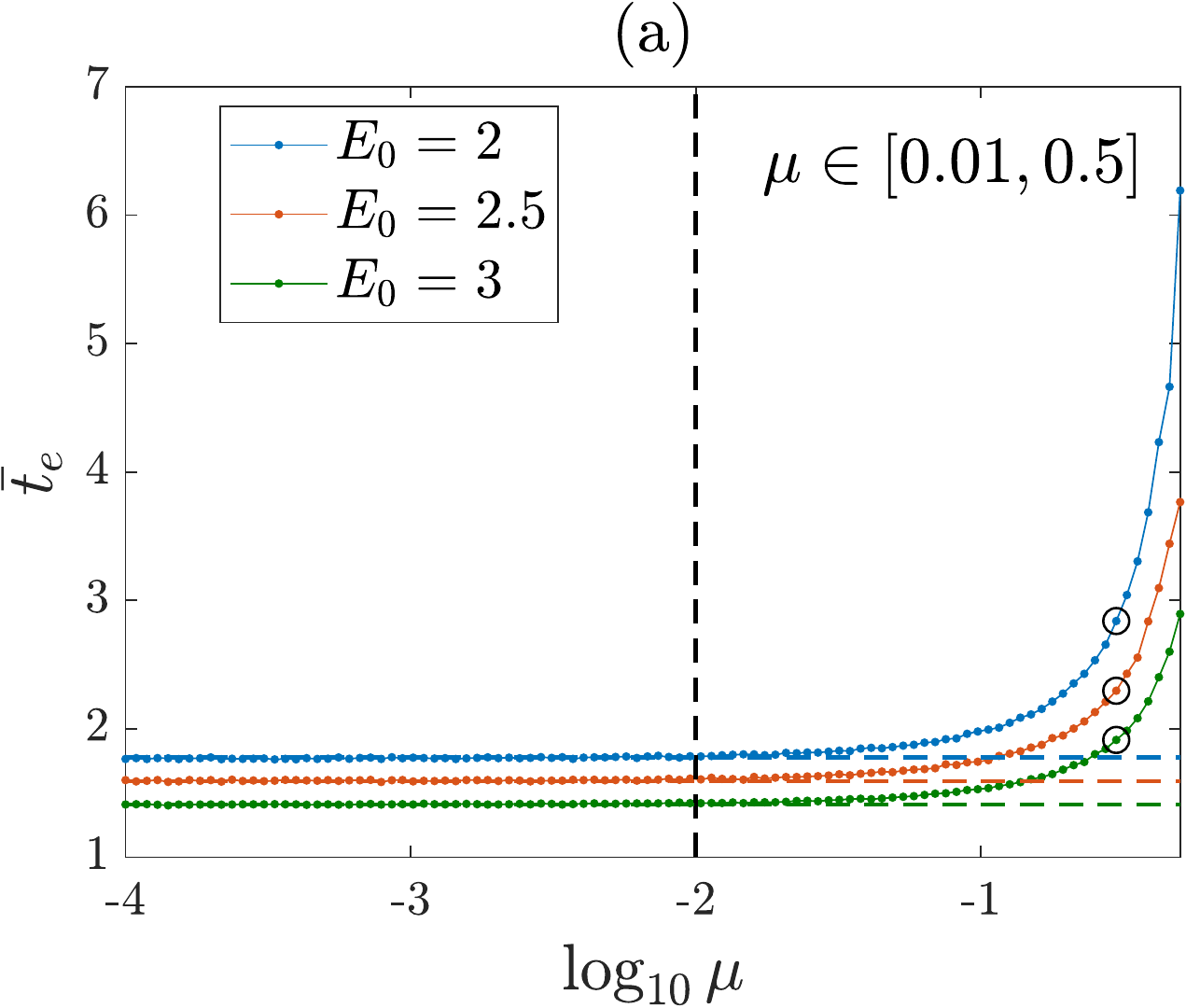}
	\quad
	\includegraphics[width=0.48\textwidth]{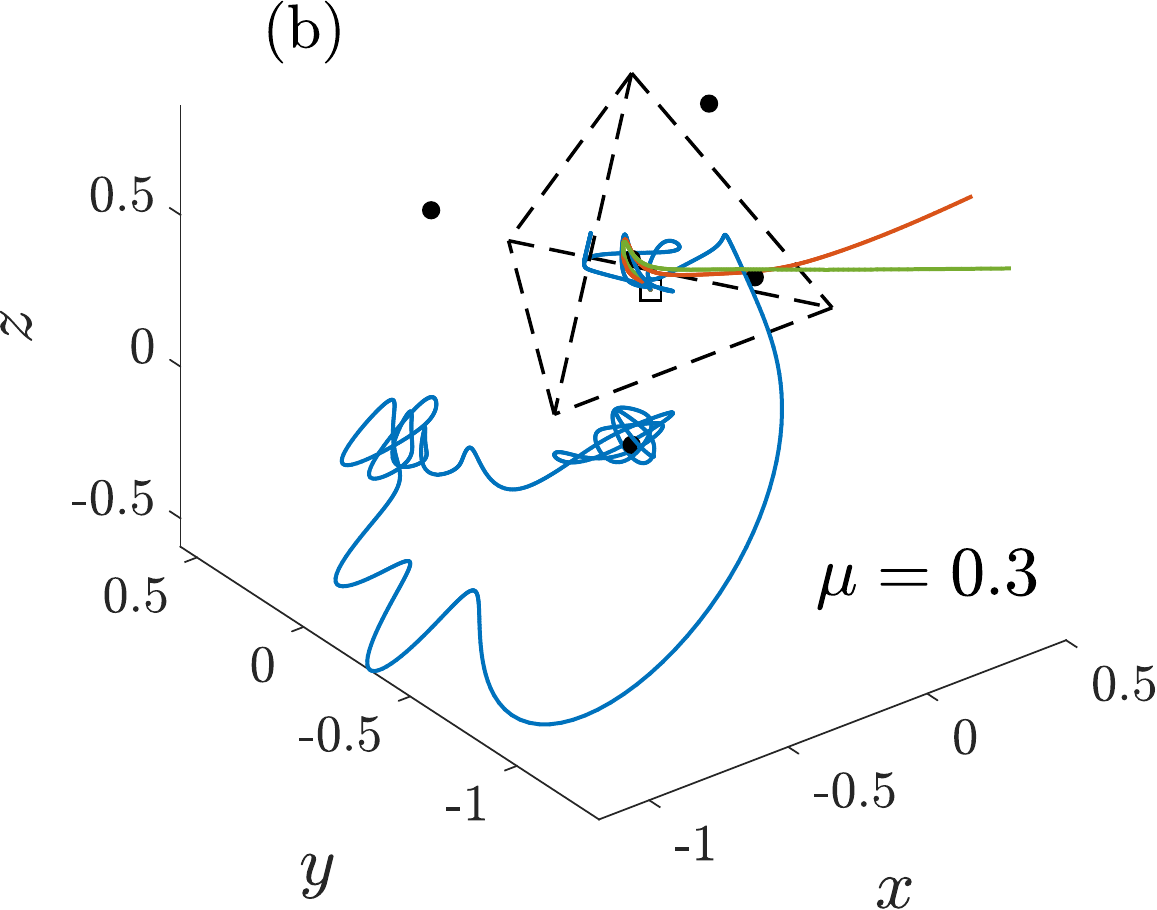}
	
	\caption{(a) Escape time average as the dissipation parameter $\mu$ is varied for fifty different values belonging to the interval $\mu \in [10^{-4},0.5]$. The three curves correspond to three initial energy cases $E_0 = 2 < E_c$ (blue), $E_0 = 2.5 > E_c$ (red) and $E_0 = 3 \gg E_c$ (green), where $E_c \approx 2.25$. The horizontal dashed lines represents the escape time average in the conservative case. The vertical dashed line is placed at $\mu = 10^{-2}$ and indicates the beginning of the regime where the dissipation significantly affects the system dynamics. The escape times have been computed by launching 2500 particles from the plane $z = 0$ toward the tetrahedron centroid and taking into account if $l > 3$. Finally, the three black circles correspond to the parameter values of the trajectories shown in (b). These trajectories are launched from the same initial condition (square) and evolve for $\mu = 0.3$ and $E_0 = 2$ (blue), $E_0 = 2.5$ (red) and $E_0 = 3$ (green), respectively. Finally, we depict the five attractors and the scattering region (black).}
	\label{fig:2}
\end{figure*}

Furthermore, here we focus on the dissipative case, in which the system initial energy $E_0$ decreases as the time passes by. As is common in this kind of problems, we introduce the dissipation parameter $\mu$ in Hamilton's equations of motion by adding a term proportional to the particle velocity in the three degrees of freedom \cite{kandrup2004}, yielding \begin{equation} \begin{aligned} \dot{x} = p_x, \quad \dot{p_x} = - \frac{\partial V}{\partial x} - \mu p_x, \\ \dot{y} = p_y, \quad \dot{p_y} = - \frac{\partial V}{\partial y} - \mu p_y, \\ \dot{z} = p_z, \quad \dot{p_z} = - \frac{\partial V}{\partial z} - \mu p_z. \end{aligned} \label{eq:5} \end{equation} Then, five attractors appear when the dissipation is introduced to which the particle may converge and be trapped forever. Specifically, one of them appears at the tetrahedron centroid, whereas the remaining four attractors are associated with potential relative minima, depicted as blue dots in Fig.~\ref{fig:1}. As four attractors lie outside the defined scattering region, we take into account in the dissipative case that a particle escapes when it moves an arbitrary distance $l$ away from the tetrahedron centroid, e.g., $l > 3$, such that it is highly unlikely that the particle can be trapped in any attractor. On the other hand, Fig.~\ref{fig:2}(a) shows how the escape time average is affected by the dissipation. We notice that, from $\mu > 10^{-2}$, the escape time averages in the dissipative and conservative cases begin to differentiate and from such a value the introduced dissipation becomes relevant. So, we use the value range $\mu \in [0.01,0.5]$ along the present work.

Three examples of trajectories are plotted in Fig.~\ref{fig:2}(b). In particular, they evolve for the same dissipation rate but depart from different values of $E_0$. Interestingly, the trajectory starting from the lowest initial energy manages to escape from the scattering region in the first term, but is eventually trapped in one of the attractors. Conversely, the other two trajectories escape quickly from the tetrahedron barely interacting with the potential. This last kind of behavior is the one we observe the most in the computations under weak dissipation.

Finally, the conservative (i.e., $\mu = 0$) and dissipative dynamics is integrated by means of an adaptive Runge-Kutta-Fehlberg method \cite{burden2015} with a relative tolerance of $10^{-12}$ for each equation system solution. The maximum value of the integration step is $5 \times 10^{-3}$, small enough to conserve the mechanical energy along the particle trajectory in the conservative case.

\newpage

\section{Non-monotonic behavior of the chaotic saddle dimension under dissipation} \label{sec:3}

\begin{figure*}[h]
	\centering
	
	\includegraphics[width=0.48\textwidth]{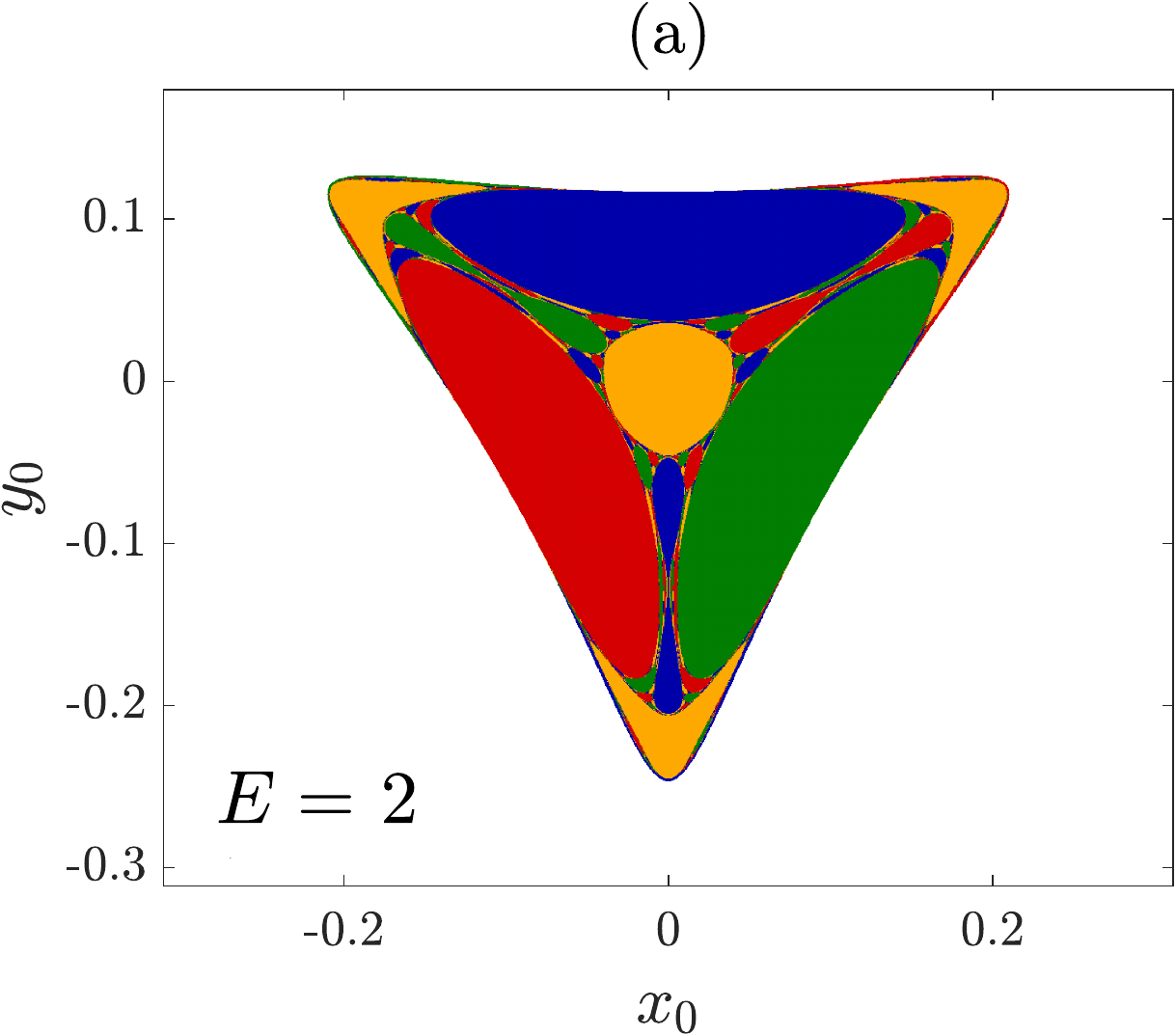}
	\quad
	\includegraphics[width=0.48\textwidth]{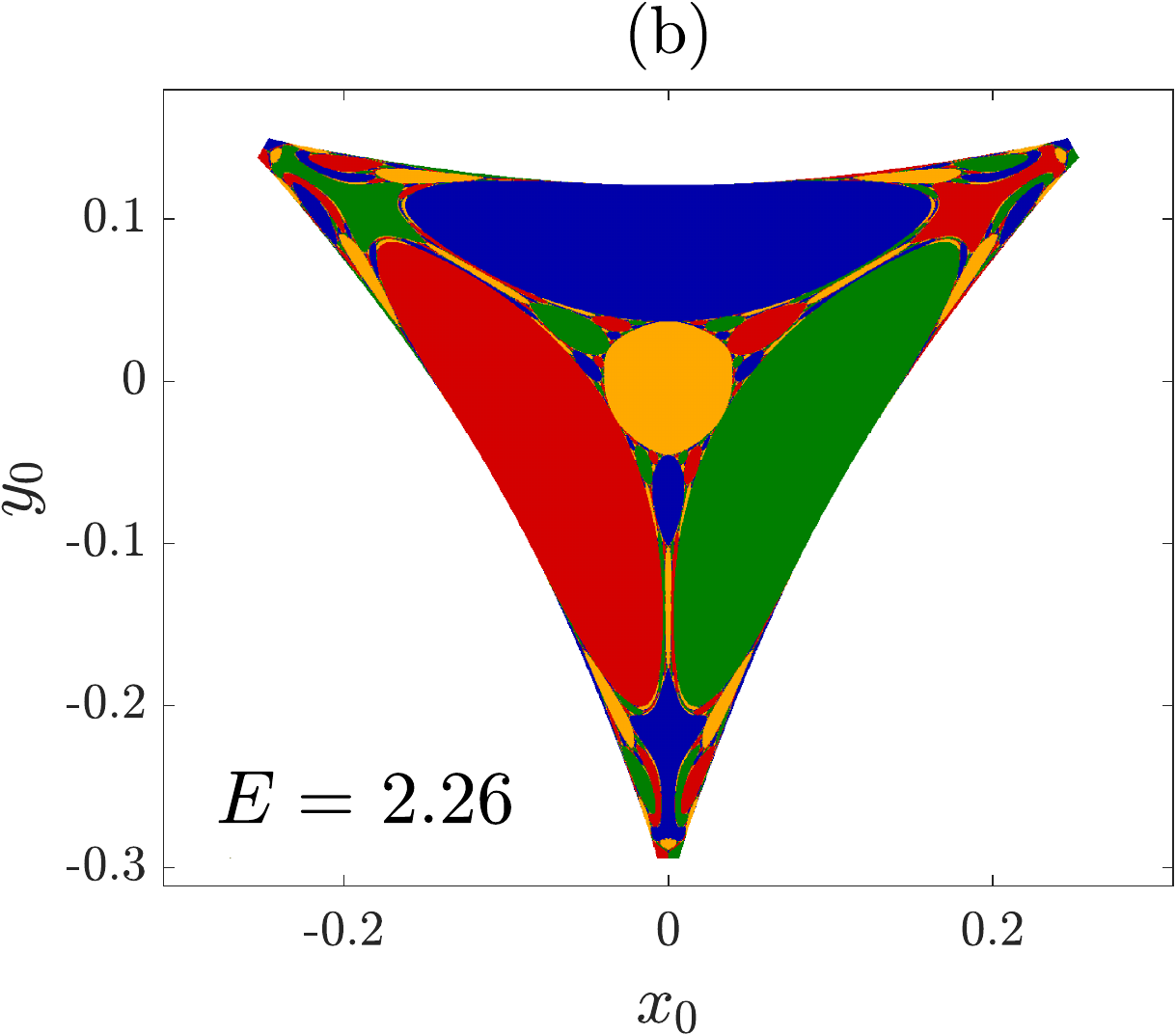}\\
	\bigskip
	\includegraphics[width=0.48\textwidth]{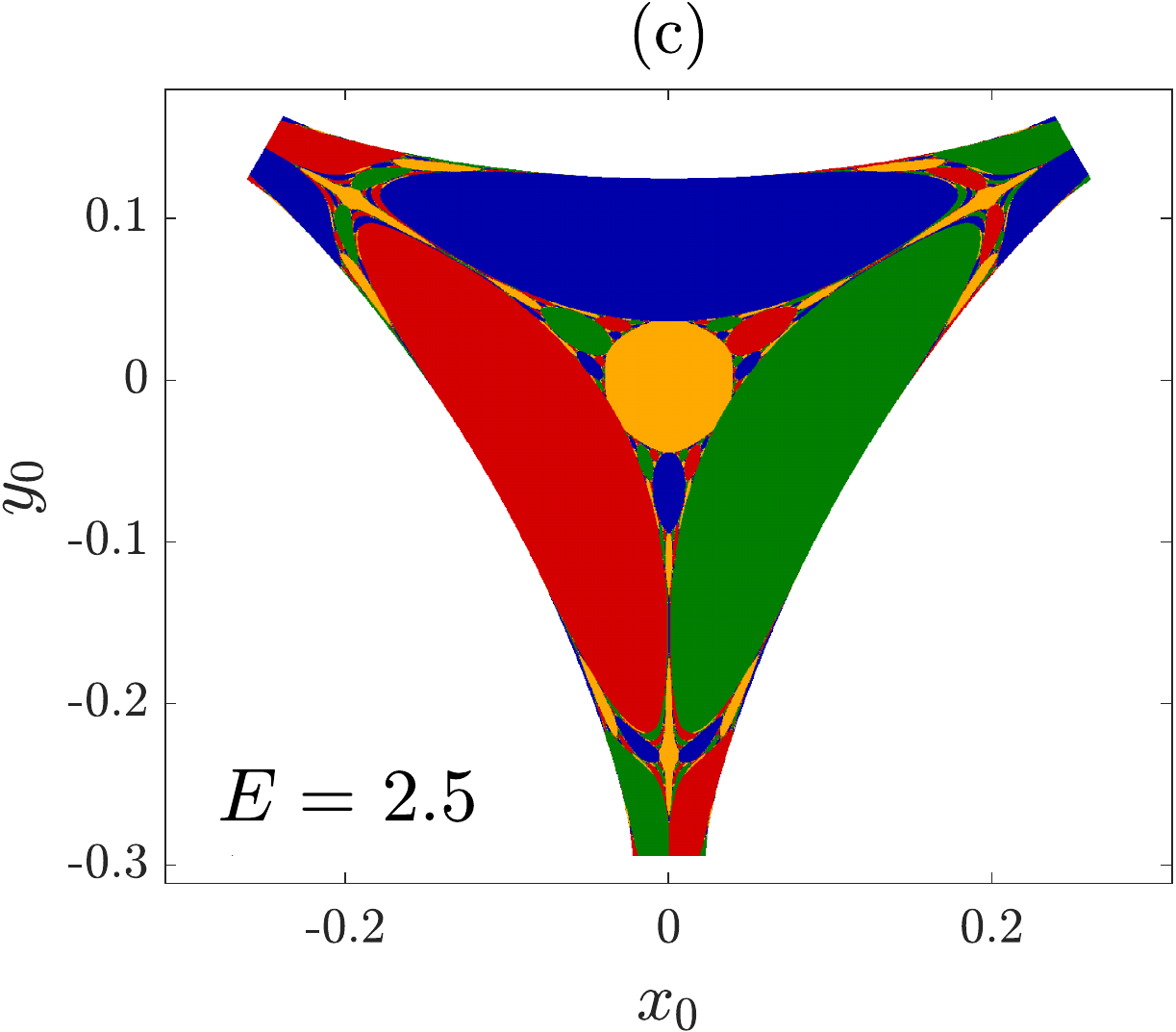}
	\quad
	\includegraphics[width=0.48\textwidth]{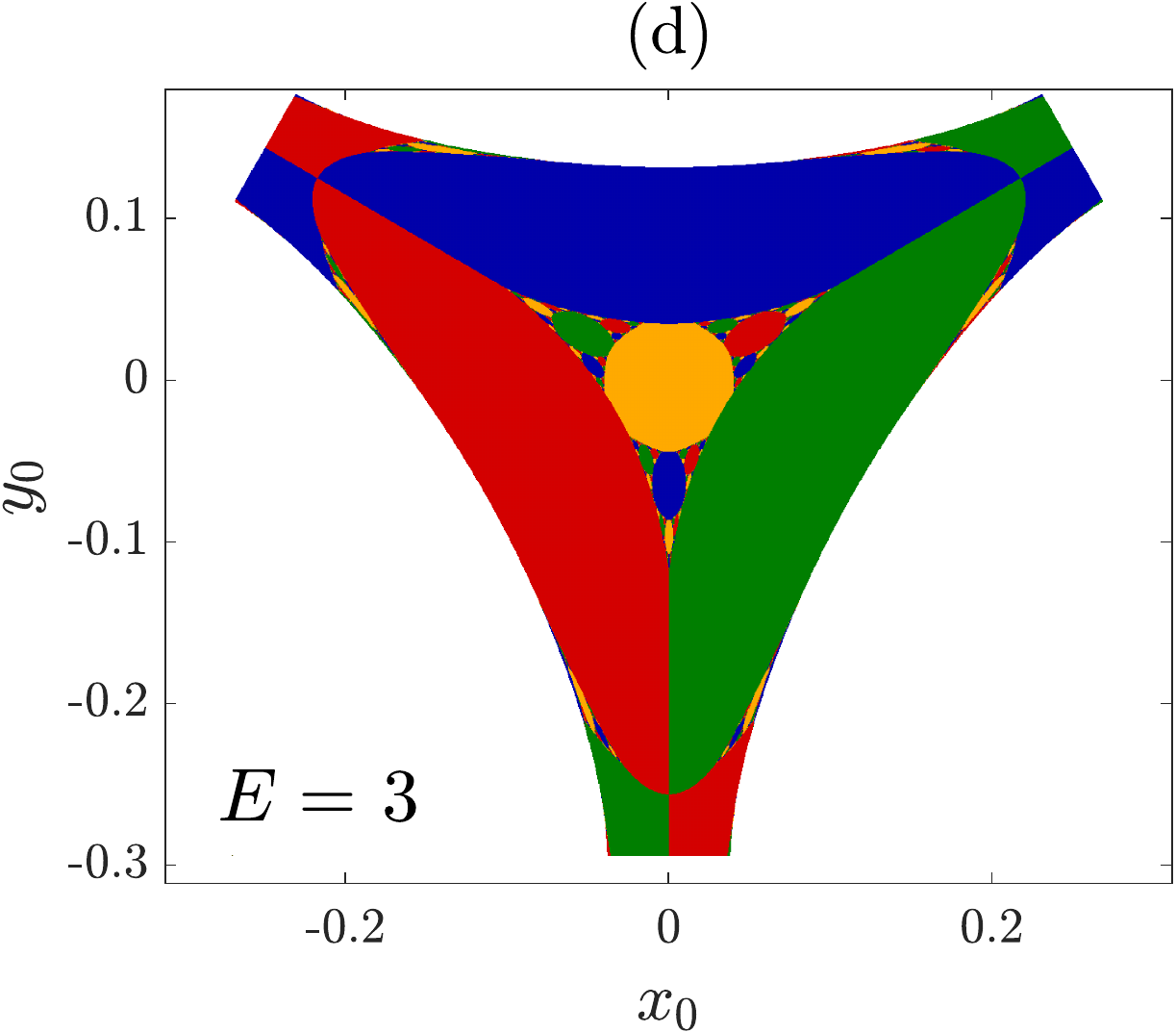}
	
	\caption{Exit basins in the conservative case for (a) $E = 2 < E_c$, (b) $E = 2.26 \approx E_c$, (c) $E = 2.5 > E_c$, and (d) $E = 3 \gg E_c$, where $E_c \approx 2.25$. The yellow color corresponds to particles that escape through the bottom plane, whereas the other colors (red, blue and green) are related to the remaining planes. Finally, the white color represents the inaccessible initial conditions, which satisfy $E < V(x,y,0)$.}
	\label{fig:3}
\end{figure*}

Along this section, we use the exit basins to analyze the system unpredictability. Specifically, we choose the plane $z = 0$ to build the basins. In addition, we launch every particle toward the tetrahedron centroid to simulate trajectories that capture all the three-dimensional scattering region phenomenology. Four basins (or colors) appear, which correspond to the four tetrahedron planes through which the particle can escape. The yellow basin is associated with the bottom exit, i.e., the plane $z = 0$, whereas the other basins (red, green and blue) are associated with the three remaining tetrahedron planes. Figure \ref{fig:3} shows the exit basins in the conservative case for several energy values. As expected, the higher the energy, the less fractal the basin boundaries. The potential function exits become wider as the energy is increased, implying the escape to happen without spending long transients. As a matter of fact, trajectories can amplify extremely small differences in their initial conditions during long transients due to the nonlinear dynamics, and eventually escape through different exits. For this reason, fractal basin boundaries are a well-known source of the unpredictability of the system final state.

The nonattracting invariant set called chaotic saddle \cite{ott1993} is responsible for the appearance of the fractal basin boundaries. The numerical computation of the fractal dimension $D_c$ of the chaotic saddle is a common method to quantify this unpredictability in chaotic scattering problems \cite{nieto2020}. For instance, the bigger the computed $D_c$ value, the more unpredictable the system, since a bigger region the phase space (the exit basins) is occupied by the chaotic saddle (the fractal basin boundaries).  In the \ref{sec:6} at the end of the manuscript we derive the expression \begin{equation} D_c = 2d + 1, \label{eq:6} \end{equation} which relates $D_c$ to the fractal dimension $d$ of the exit basin boundaries. Thus, we can determine $D_c$ after calculating $d$ with the help of the uncertainty exponent algorithm \cite{mcdonald1985} (see the \ref{sec:6} for further computation details).

\begin{figure*}[h]
	\centering
	
	\includegraphics[width=0.48\textwidth]{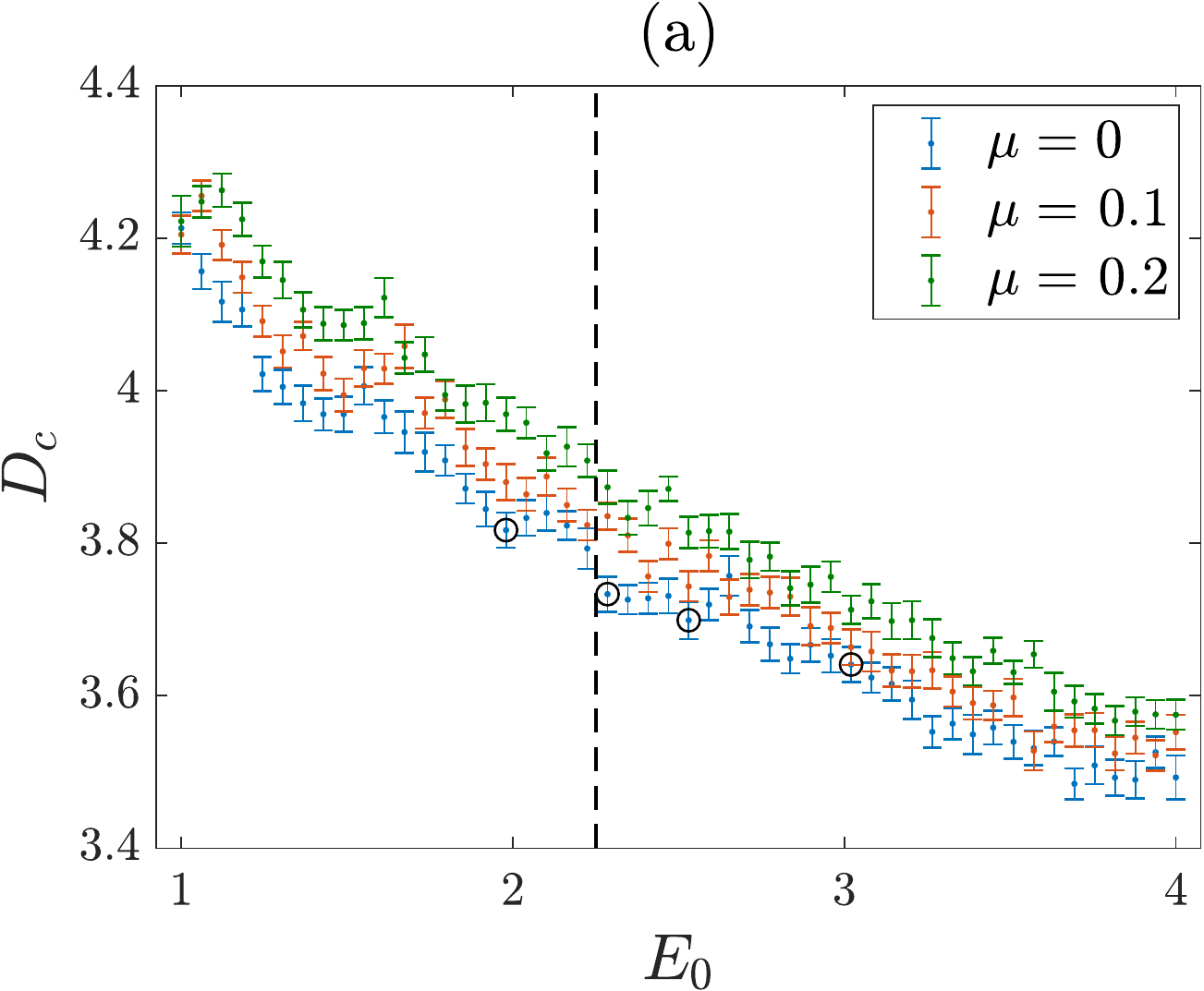}
	\quad
	\includegraphics[width=0.48\textwidth]{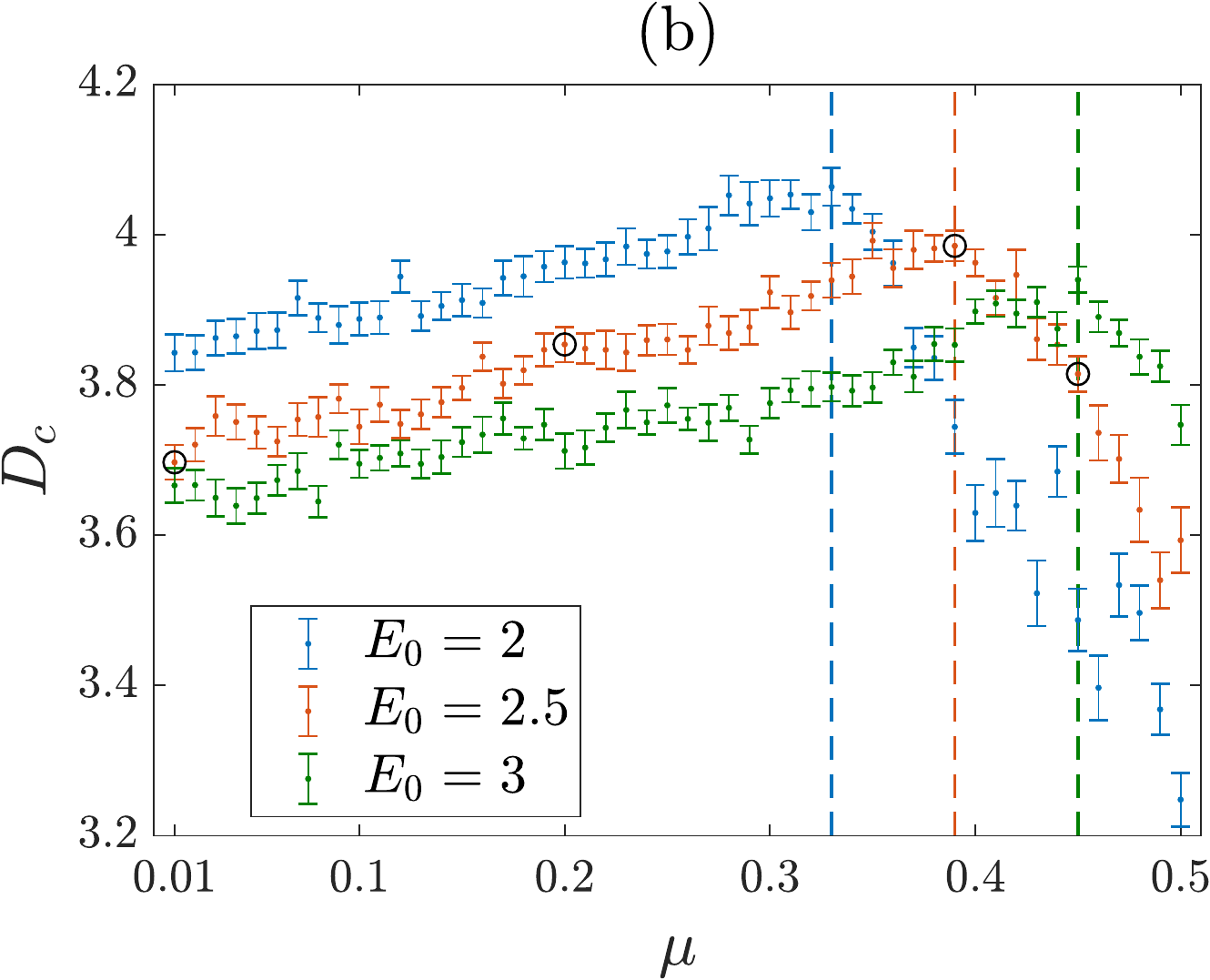}
	
	\caption{(a) Fractal dimension of the chaotic saddle as $E_0$ is varied for three different dissipation values. The vertical dashed line is placed at $E_c \approx 2.25$, and four circles are drawn at the energy values of the exit basins already shown in Fig.~\ref{fig:3}. (b) Similarly, the fractal dimension of the chaotic saddle as $\mu$ is varied for three different initial energies. The vertical dashed lines are placed at $\mu_m (E_0)$, i.e., the dissipation parameter values where the fractal dimension $D_c$ reaches its maximum for a given initial energy. Specifically, $\mu_m(2) = 0.33$, $\mu_m(2.5) = 0.39$ and $\mu_m(3) = 0.45$. Again, we draw four circles at the $\mu$ values of the exit basins shown in Fig.~\ref{fig:5}. See the \ref{sec:6} for more computation details.}
	\label{fig:4}
\end{figure*}

Therefore, we display the fractal dimension of the chaotic saddle versus the initial energy and the dissipation in Figs.~\ref{fig:4}(a) and \ref{fig:4}(b), respectively. On the one hand, the greater the initial energy, the smaller the value of $D_c$. As already mentioned, increasing the energy softens the basin boundaries, making the system to be more predictable. On the other hand, the dissipative case is more complicate. When the dissipation is weak, the fractal dimension $D_c$ increases until a maximum value of $D_c$ is obtained, as observed for three different initial energies in Fig.~\ref{fig:4}(b). Immediately after reaching this maximum value, the system unpredictability exhibits an abrupt decreasing to presumably $D_c = 3$, as theoretically reasoned in the \ref{sec:6}. This tendency of $D_c$ can be interpreted by visualizing the dissipation parameter impact on the basin boundaries.

\begin{figure*}[h]
	\centering
	
	\includegraphics[width=0.48\textwidth]{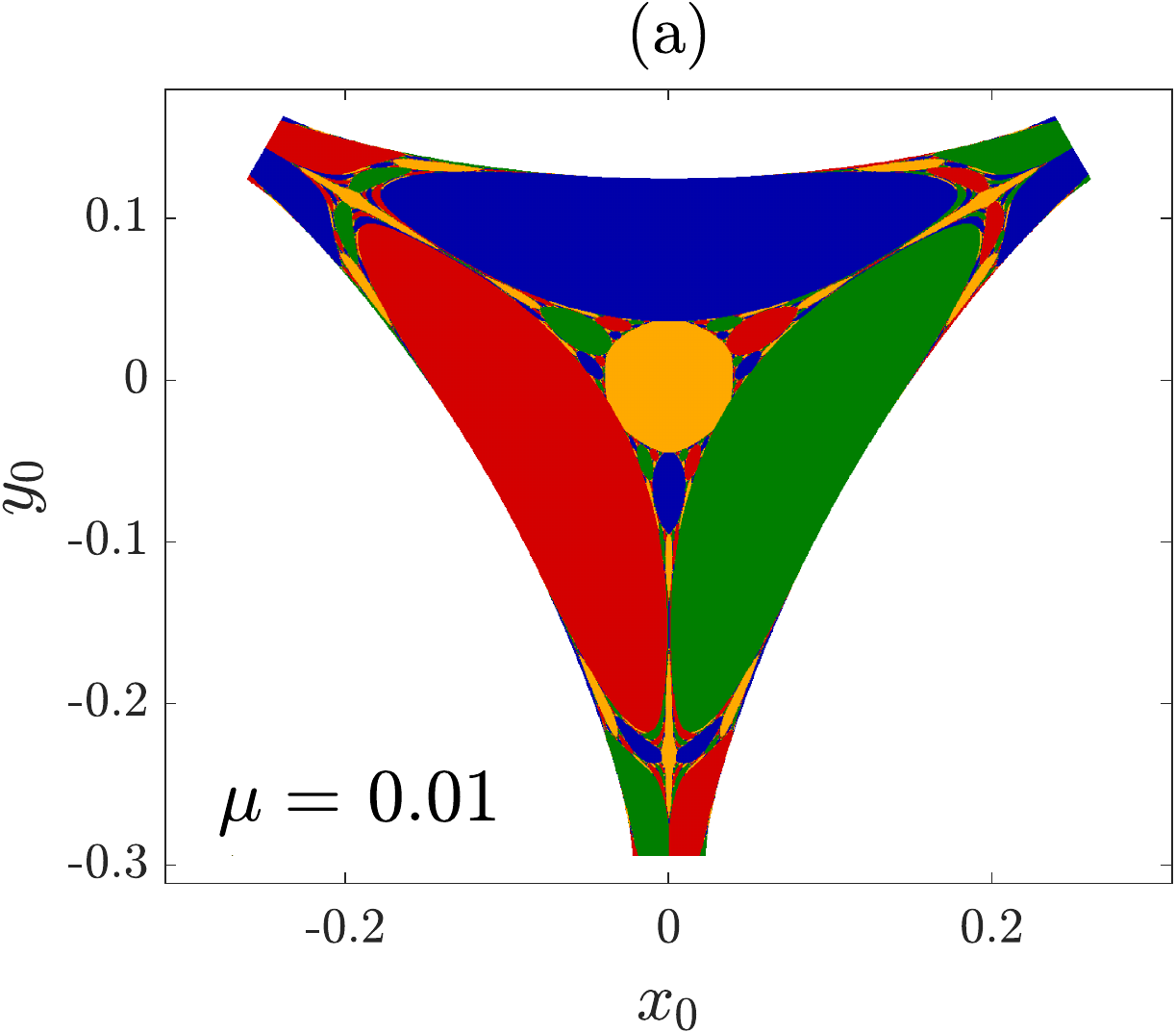}
	\quad
	\includegraphics[width=0.48\textwidth]{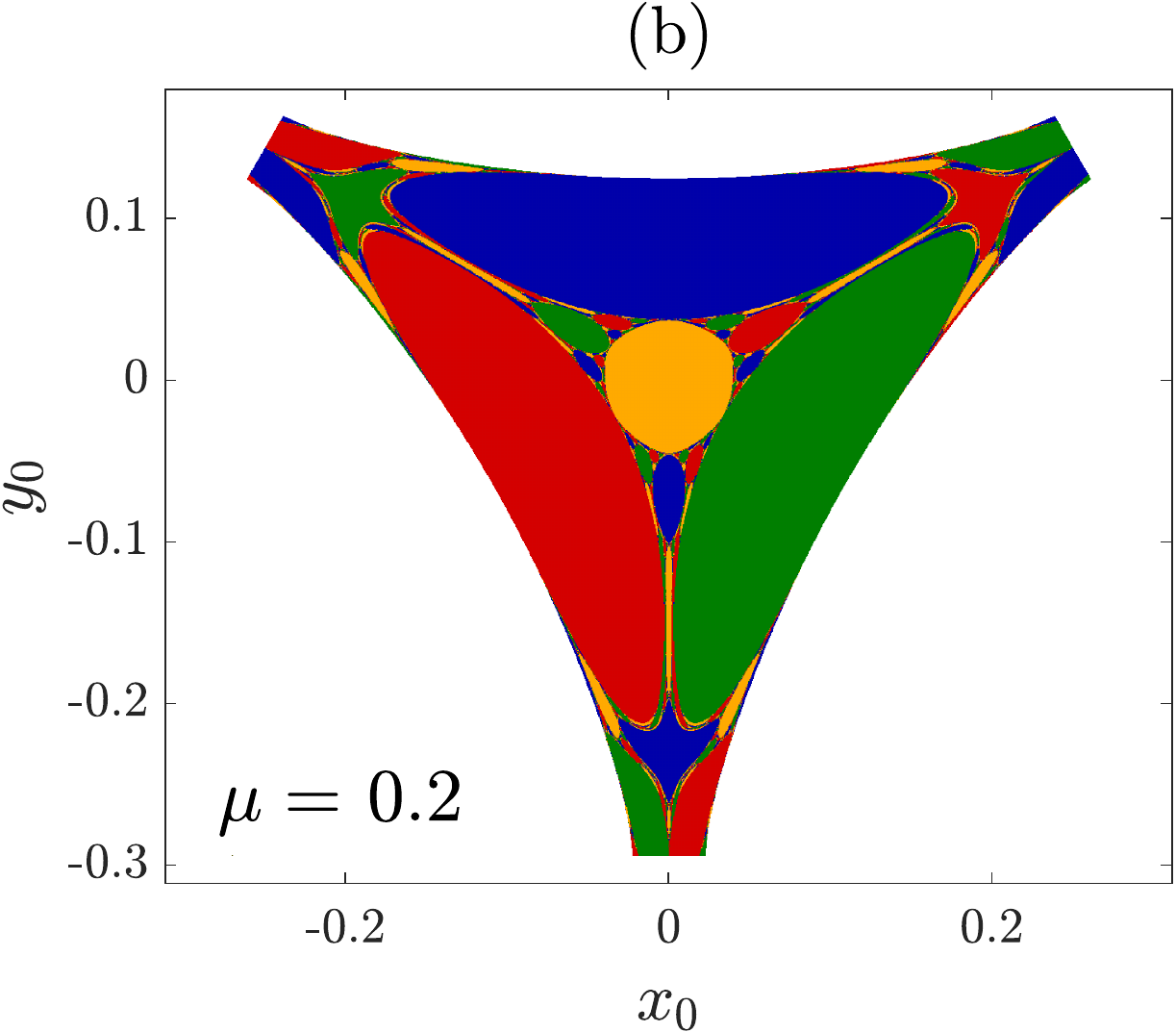}\\
	\bigskip
	\includegraphics[width=0.48\textwidth]{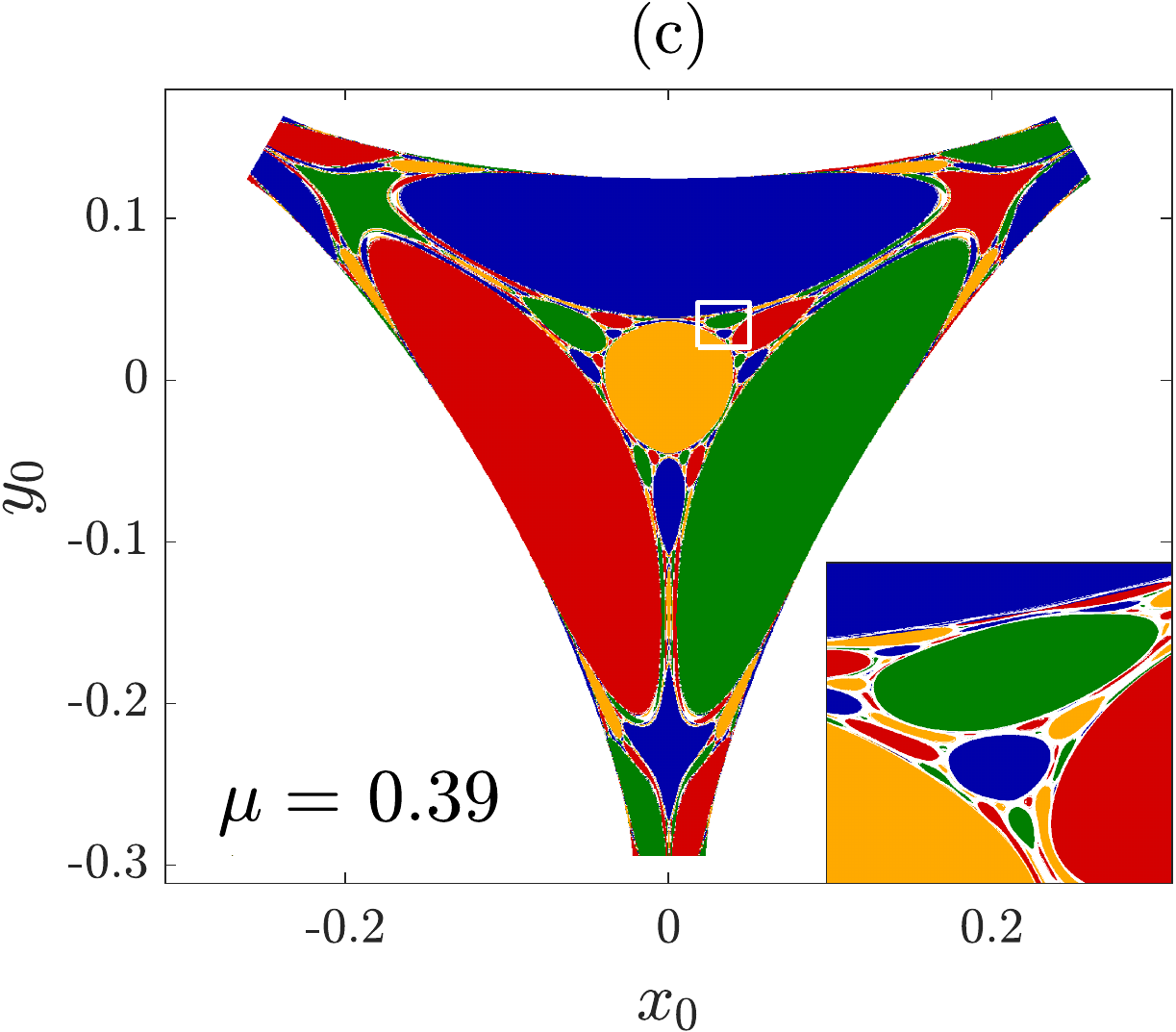}
	\quad
	\includegraphics[width=0.48\textwidth]{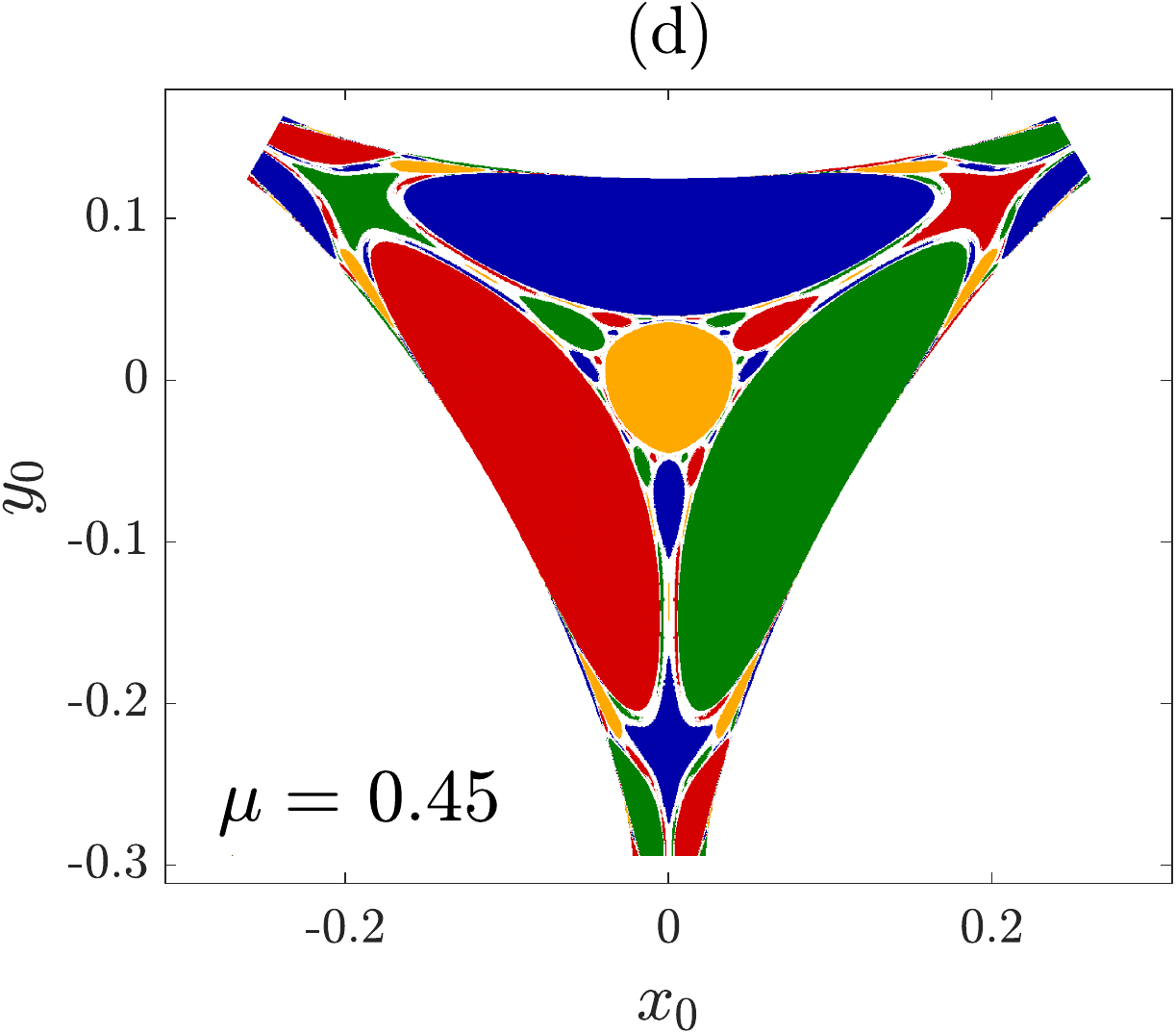}
	
	\caption{Exit basins in the dissipative case for $E_0 = 2.5$ and (a) $\mu = 0.01 \ll \mu_m$, (b) $\mu = 0.2 < \mu_m$, (c) $\mu_m = 0.39$, and (d) $\mu = 0.45 > \mu_m$. Specifically, $\mu_m = 0.39$ represents the dissipation value for which the fractal dimension is maximum for $E_0 = 2.5$. (Inset plot) We magnify a basin region to illustrate how the fractal basin boundary at $\mu_m = 0.39$ is replaced by a new white basin, which represents the initial conditions that lead trajectories to be trapped in the attractors.}
	\label{fig:5}
\end{figure*}

Hence, Fig.~\ref{fig:5} shows four dissipative cases for $E_0 = 2.5$. Firstly, the basins are barely affected by the dissipation for $\mu = 0.01$. However, the higher the dissipation, the longer the transients within the scattering region. Then, as observed for $\mu = 0.2$, the basin boundaries widen and the value of $D_c$ increases. A weak dissipation makes the system to be more unpredictable. In addition, as the dissipation is increased, some trajectories starting from the basin boundaries can be trapped. Then, another basin (or color) appears and the basins become more complex and unpredictable. In this situation, the fractal dimension $D_c$ grows until a maximum value is reached when all trajectories starting from the boundaries are trapped. The latter is satisfied for $\mu = 0.39$, where the fractal basin boundaries are completely erased by a new white and smooth basin. Here we can define $\mu_m$ as the dissipation parameter value that causes the fractal dimension $D_c$ to be maximum. In particular, the value of $\mu_m$ depends on the initial energy, since the larger $E_0$ the more dissipation must be introduced to trap particles. Finally, the new white basin grows when the dissipation is stronger than $\mu_m$, as seen for $\mu = 0.45$. Therefore, the fractal dimension $D_c$ sharply declines in this scenario of strong dissipation.

\section{Driving and enhancement of Wada basins under dissipation} \label{sec:4}

Now we investigate how dissipation affects the topology of the basin boundaries, which may exhibit the Wada property. Briefly, Wada basins share a common boundary implying that between two basins it is always possible to find all other basins. The exit basin boundary is non Wada for $E>E_c$ in the conservative regime. That is, only some isolated points on the boundaries are Wada points, which are defined as points separating the four basins (or colors).  For $E>E_c$, the inaccessible regions remain disconnected, causing a part of the tetrahedron edges to be accessible. The exits, i.e., the tetrahedron planes, are connected by the part of the edges not swallowed by the inaccessible regions. Hence, it is possible to propose a continuous and smooth path of initial conditions whose corresponding trajectories escape through an exit at the path beginning and through another different exit at the path ending. Then, necessarily, a trajectory must escape at some path point through the tetrahedron edge separating the two exits. This hypothetical path of initial conditions consists of two one-dimensional basins separated by a boundary where only two colors meet, i.e., a non-Wada point.

On the other hand, the exit basin boundary exhibits the Wada property for $E < E_c$, i.e., all boundary points are Wada points. The inaccessible region is connected and completely covers the tetrahedron edges. Hence, the exits remain disconnected. In this case, there is no hypothetical path of initial conditions whose corresponding trajectories escape through only two exits. These trajectories, as they approach the now inaccessible edge of the tetrahedron, deviate until escaping inevitably through a third exit different from the previous ones. Note that this last argument is valid for any path separating two basins, no matter its length. Therefore, a different basin always appears between any two basins, and so on. For $E < E_c$, the path is formed by two one-dimensional basins separated by a Wada boundary, consisting of an infinite number of Wada points.

\begin{figure*}[h]
	\centering
	
	\includegraphics[width=0.5\textwidth]{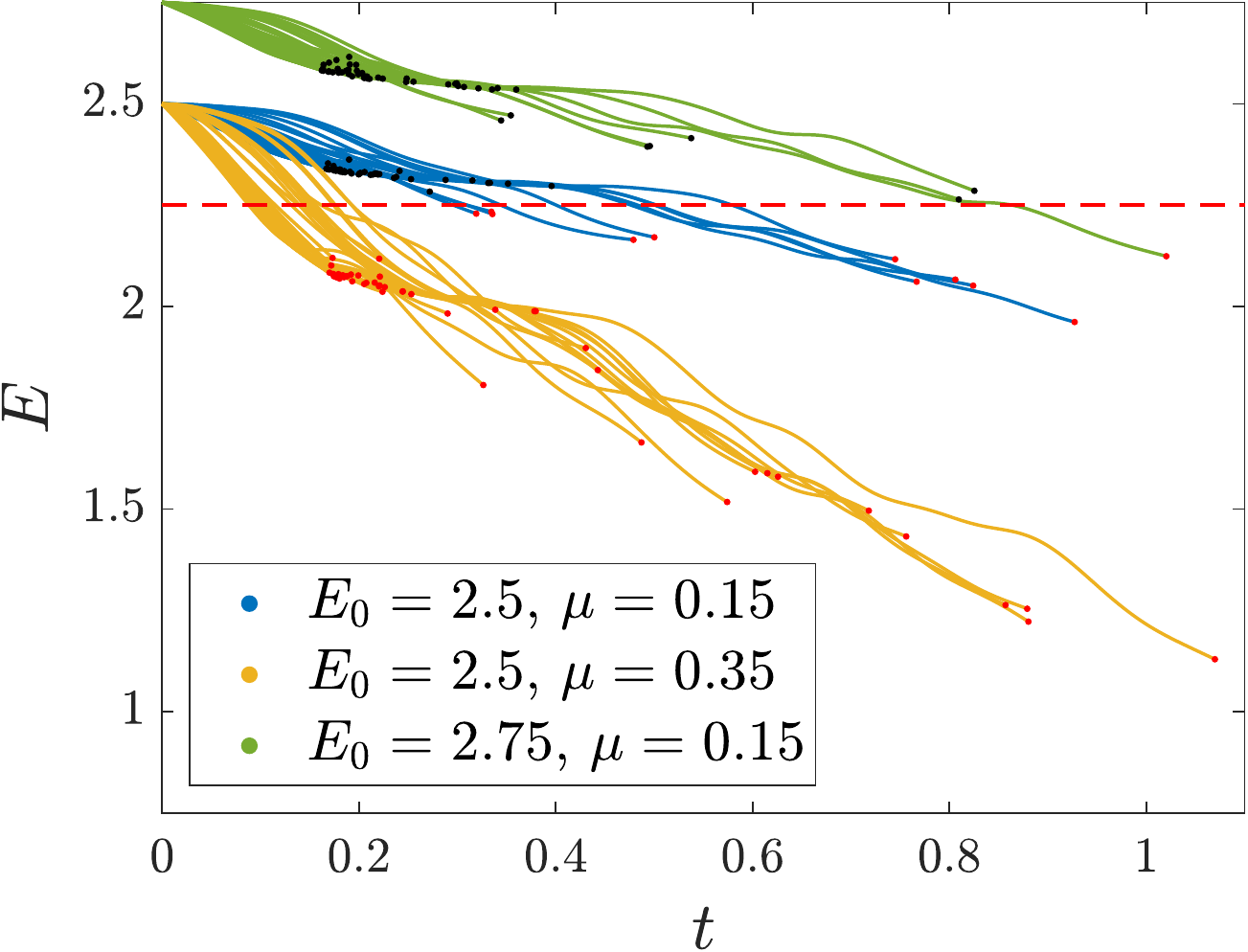}
	
	\caption{Energy values of fifty particles over time for three dissipative cases, namely, $E_0 = 2.5$ and $\mu = 0.15$ (blue), $E_0 = 2.5$ and $\mu = 0.35$ (yellow), and $E_0 = 2.75$ and $\mu = 0.15$ (green). We draw a red (black) dot if the particle final energy before leaving the tetrahedron is smaller (greater) than the critical energy $E_c \approx 2.25$, which is indicated by the horizontal red dashed line.}
	\label{fig:6}
\end{figure*}

In this context, where the system energy determines whether the basins are Wada or not, introducing energy dissipation may affect their topology. For instance, as illustrated in Fig.~\ref{fig:6}, a particle starting from $E_0 > E_c$ may leave the tetrahedron with $E< E_c$ due to dissipation after spending a finite time within the scattering region. Hence, this particle escapes from the tetrahedron with the inaccessible regions fully connected. Regarding the latter, it is worth recalling that a dissipative system can be understood as a sequence of momentarily conservative systems of decreasing energy over time. So, the momentarily inaccessible regions of the potential function grow in radius as the energy is dissipated. Then, when all particles starting from the basin boundaries escape with energy values lower than the critical energy, only Wada paths of initial conditions are possible. In this manner, from a theoretical point of view, non-Wada basins expected in the conservative case may become Wada basins under dissipation. Nonetheless, when the dissipation allows only a few particles to escape with energies below the critical energy, the basin boundaries are formed by Wada and non-Wada points, which is known as the partially Wada property \cite{zhang2013}. In this situation, only trajectories with long transients contribute to the formation of Wada points. Furthermore, as detailed in the previous section, a strong dissipation is expected to reduce both the fractal boundary and the number of Wada points.

Now we test the emergence of the Wada points in the dissipative case by applying a recently developed numerical method called \textit{merging method} \cite{daza2018}. The method is based on a non-obvious aspect of Wada basins: if the boundary of a basin is Wada, then this boundary remains invariant under the merging of the other basins. Thus, the method consists of verifying this latter statement by merging basins. On the other hand, an advantage of this method is that it only requires a basin image of finite resolution as input. This is useful when there is no more knowledge about the system and its dynamics than a basin image. However, the method has also limits, since it only verifies whether a boundary point is Wada or non-Wada up to a given resolution. In addition, the Wada point detection depends strongly on the fattening parameter $r$. The method only ascertains what the fraction of Wada points is given a value of $r$. In practice, the computation has to be repeated increasing $r \in \mathbf{N}$ to test how the fraction of Wada points evolves as a function of $r$. Finally, the computation ends at the value of $r$ for which the Wada point fraction as a function of $r$ has converged. The steps and many other aspects of this method are extensively detailed in Ref.~\cite{daza2018}.

\begin{figure*}[h]
	\centering
	
	\includegraphics[width=0.485\textwidth]{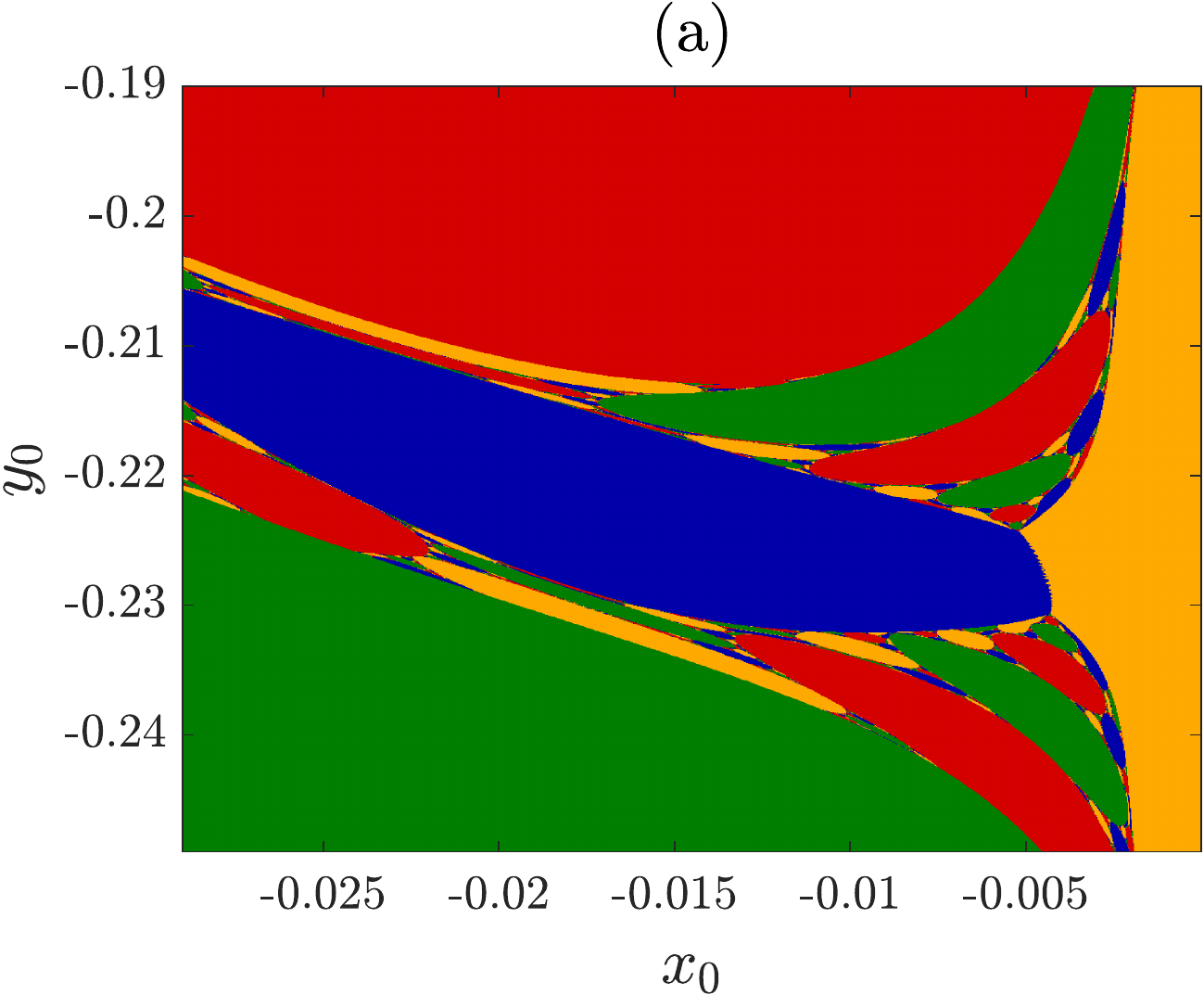}
	\quad
	\includegraphics[width=0.46375\textwidth]{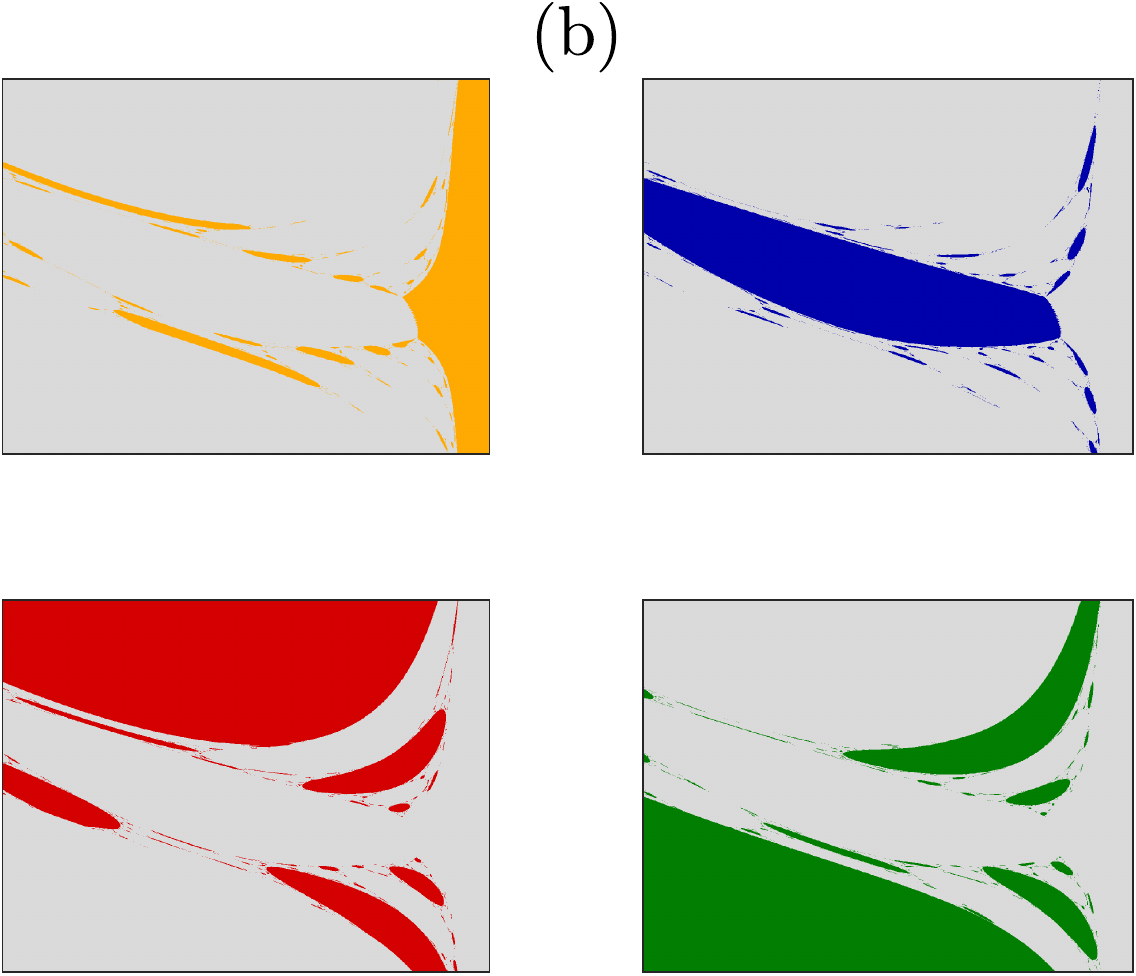}
	
	\caption{(a) Exit basins as computed for $E_0 = 2.5$ and $\mu = 0.01$ with a 1000$\times$1000 resolution. We focus on the magnified region $x_0 \in [-0.029, 0)$ and $y_0 \in [-0.249, -0.19]$, because non-Wada points can be detected when considering these parameter values and finite resolution. Then, here we implement the \textit{merging method} to compute the Wada point fraction as the dissipation parameter is varied \cite{daza2018}. (b) The four merged basins are represented in gray and are obtained by merging the remaining basins (including the white basin) except the colored one, namely, yellow, blue, red and green, respectively.}
	\label{fig:7}
	
\end{figure*}

We provide an example of merged basins for $E_0 = 2.5$ in Fig.~\ref{fig:7}. Here we only focus on a convenient basin region where non-Wada points can be detected when considering a finite resolution for $E_0 = 2.5$. As a matter of fact, the vast majority of boundary points seem to be Wada at lower resolutions despite $E>E_c$ (e.g., see again Fig.~\ref{fig:3}(c)). We have verified by successive magnifications of the basin boundary that, at higher resolution, the boundary points only separate two basins for $E > E_c$ in the conservative case, except for some isolated Wada points where the four colors meet. This kind of fractal structure is known as a slim fractal \cite{chen2017}, which is simpler at smaller resolutions.

\begin{figure*}[h]
	\centering
	
	\includegraphics[width=0.48\textwidth]{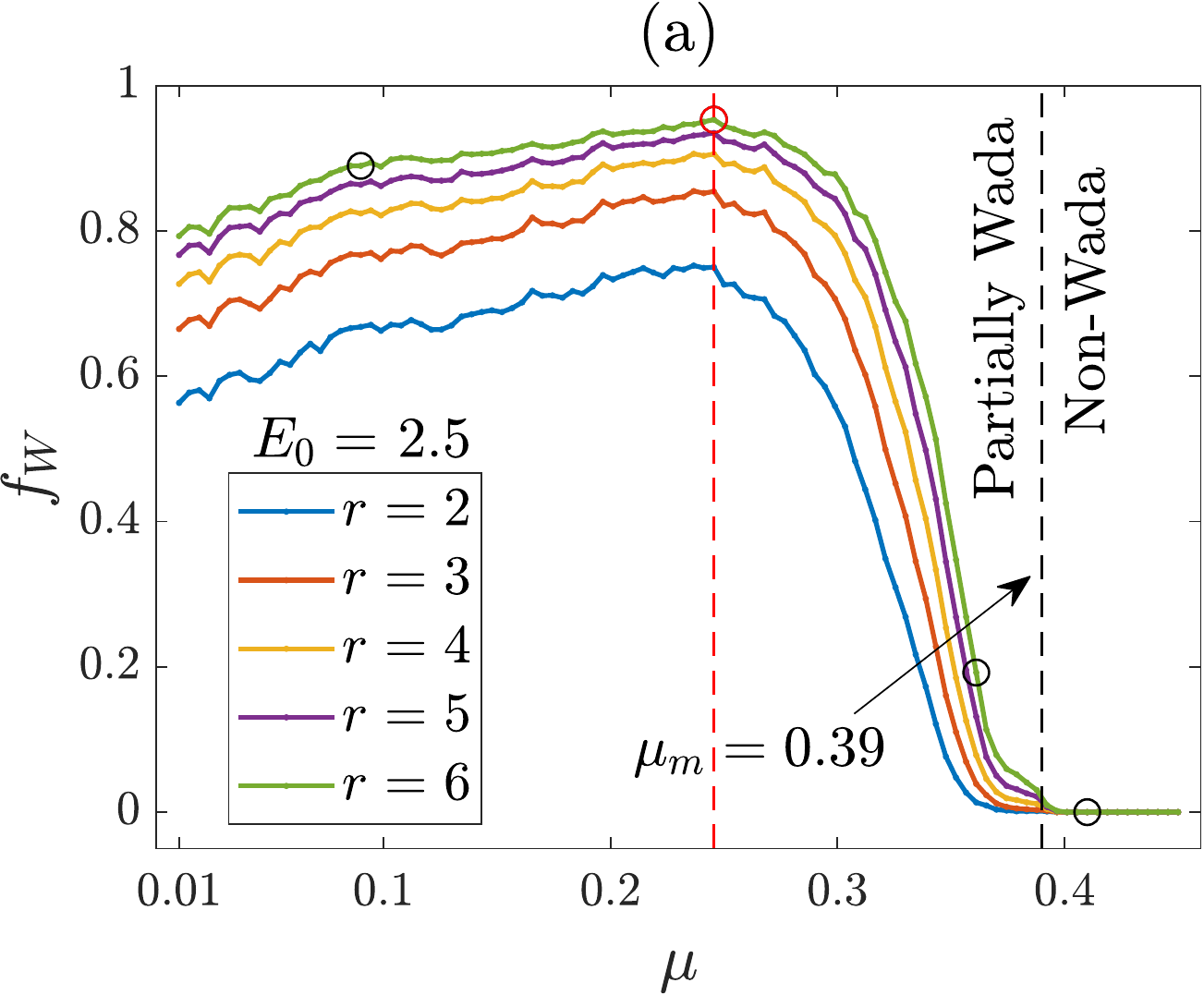}
	\quad
	\includegraphics[width=0.48\textwidth]{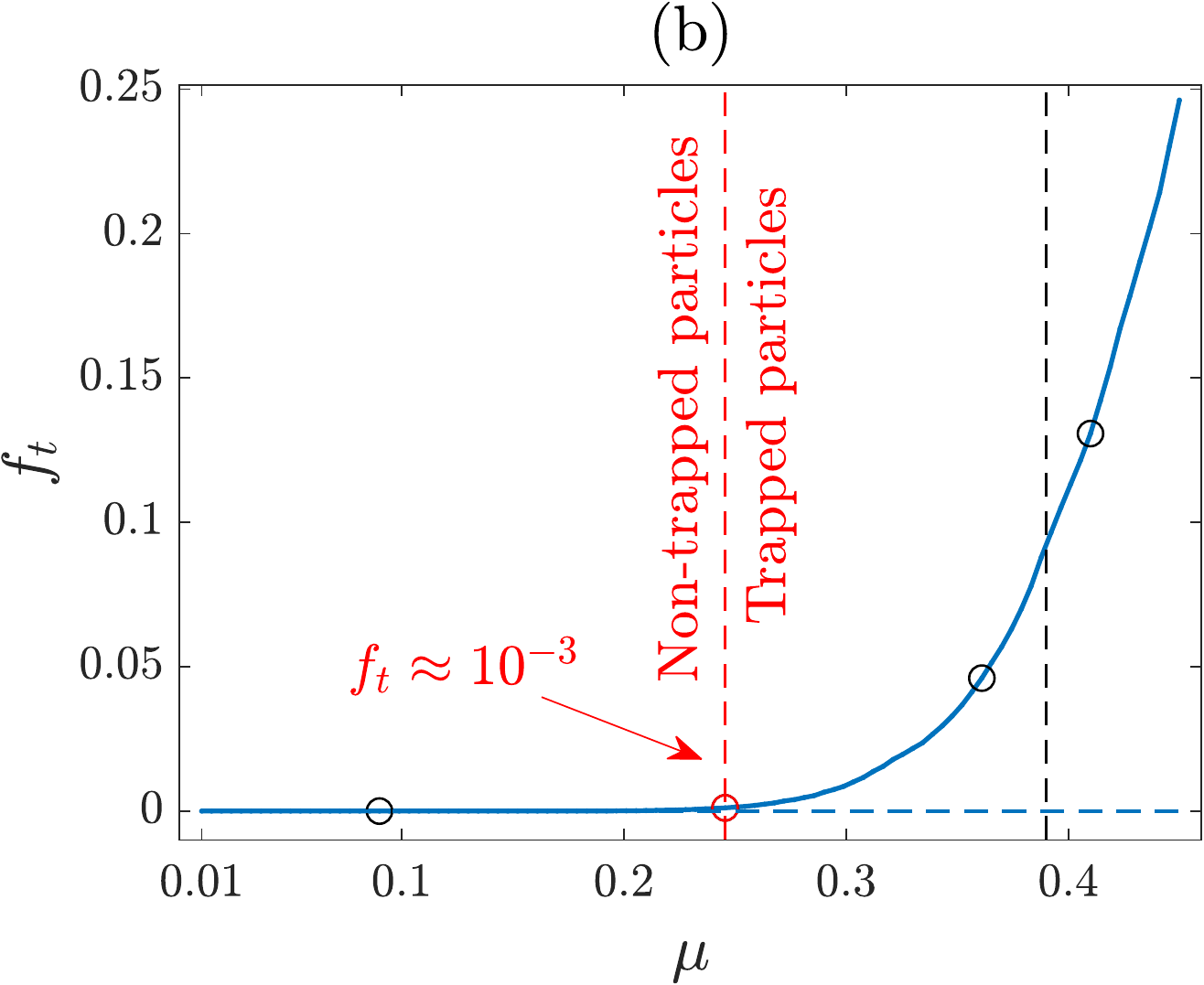}
	
	\caption{(a) The fraction of Wada points $f_W$ on the basin boundaries for several values of the fattening parameter and (b) the fraction of trapped particles $f_t$ both versus one hundred values of the dissipation parameter belonging to the interval $ \mu \in [0.01, 0.45]$ and $E_0 = 2.5$. We draw a red dashed line at the maximum value of $f_W$, which is reached when dissipation begins to be strong enough to trap particles, as indicated in (b) as well. On the other hand, $\mu_m = 0.39$  (black dashed line) represents strong dissipation for $E_0 = 2.5$ and, regarding the basin boundary topology, it indicates the separation between the partially Wada \cite{zhang2013} and non-Wada regimes. Furthermore, we draw four circles at the values of $\mu$ of the basins shown in Fig.~\ref{fig:9}.}
	\label{fig:8}
\end{figure*}

\begin{figure*}[htp!]
	\centering
	
	\includegraphics[width=0.48\textwidth]{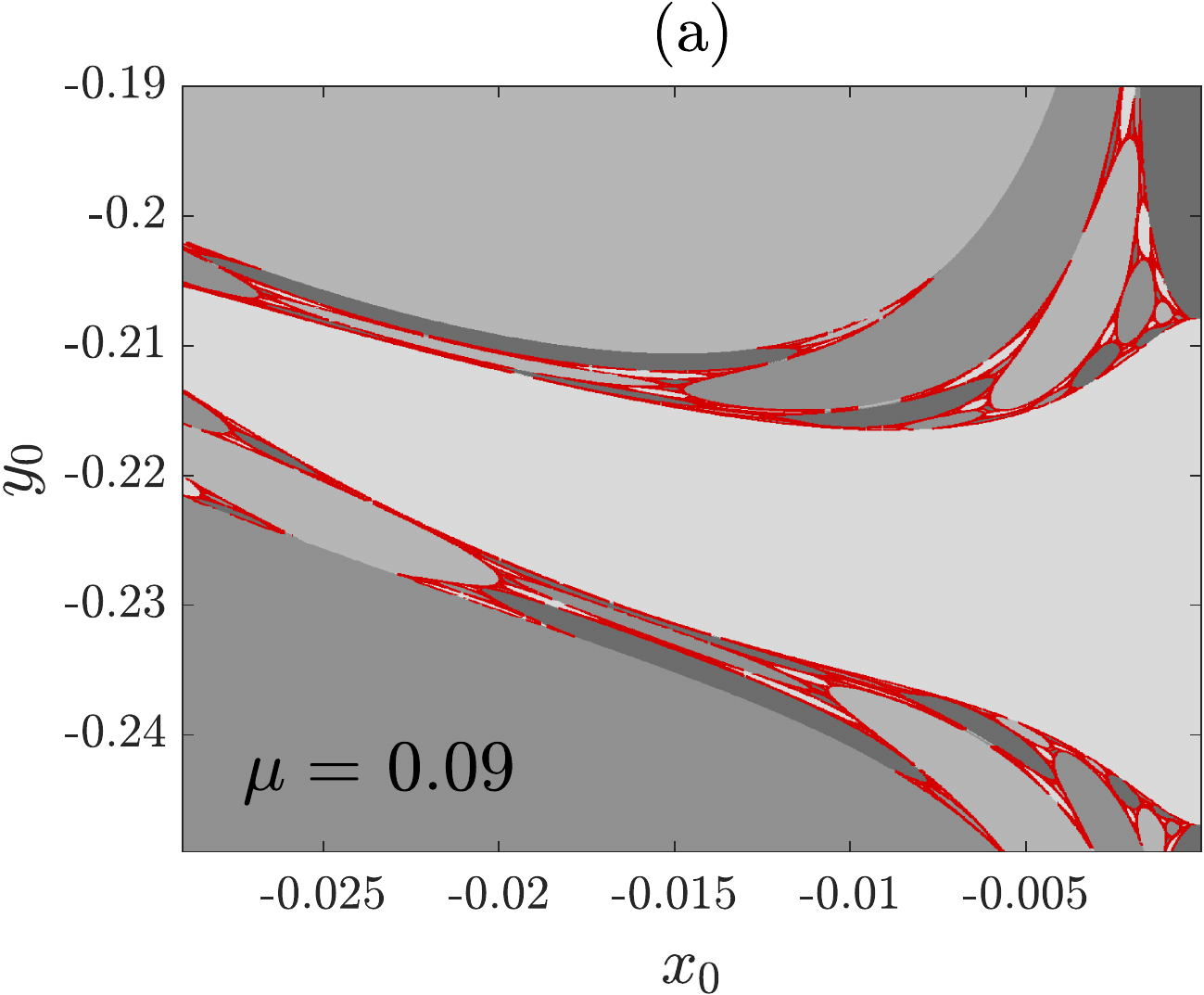}
	\quad
	\includegraphics[width=0.48\textwidth]{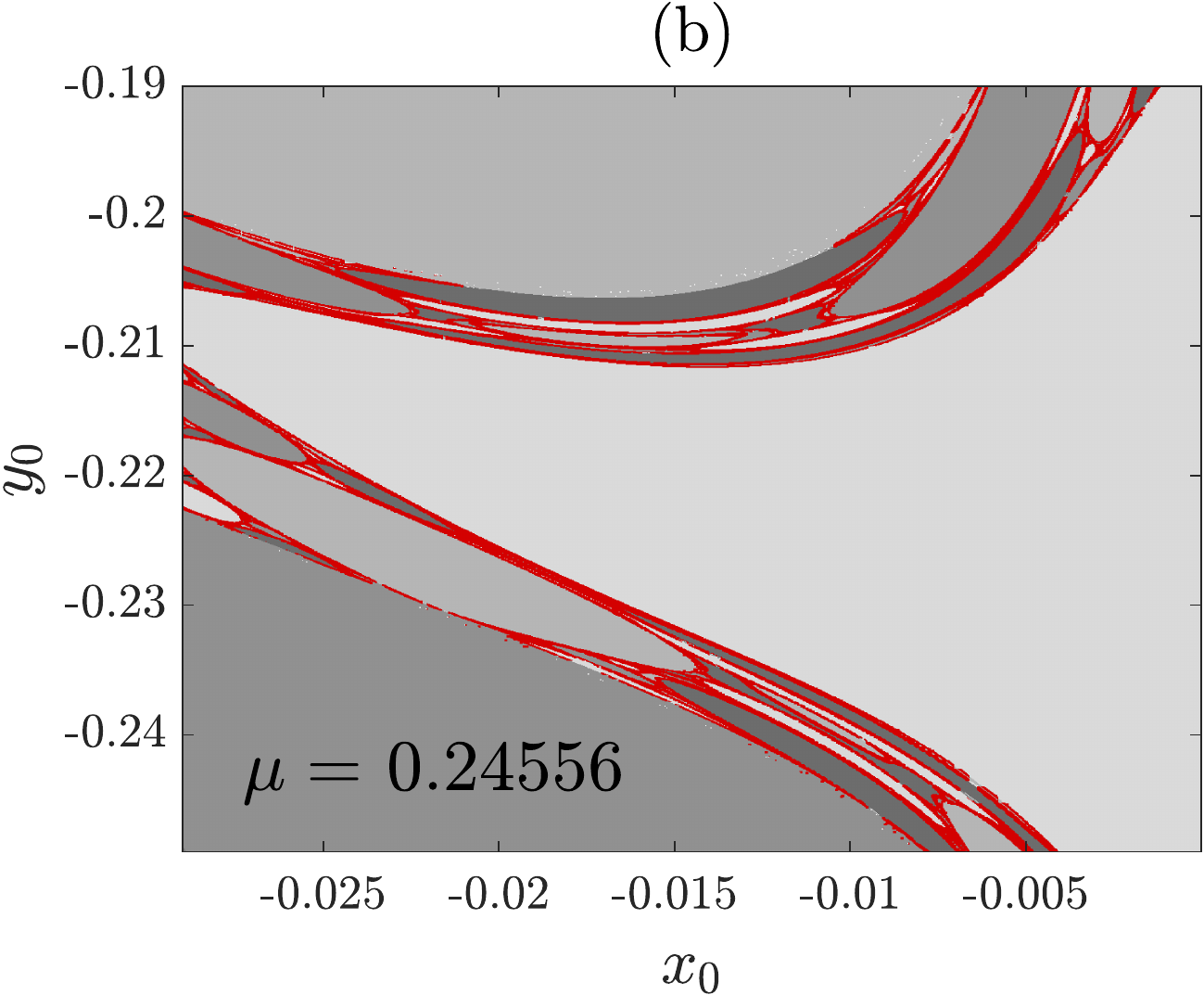}\\
	\bigskip
	\includegraphics[width=0.48\textwidth]{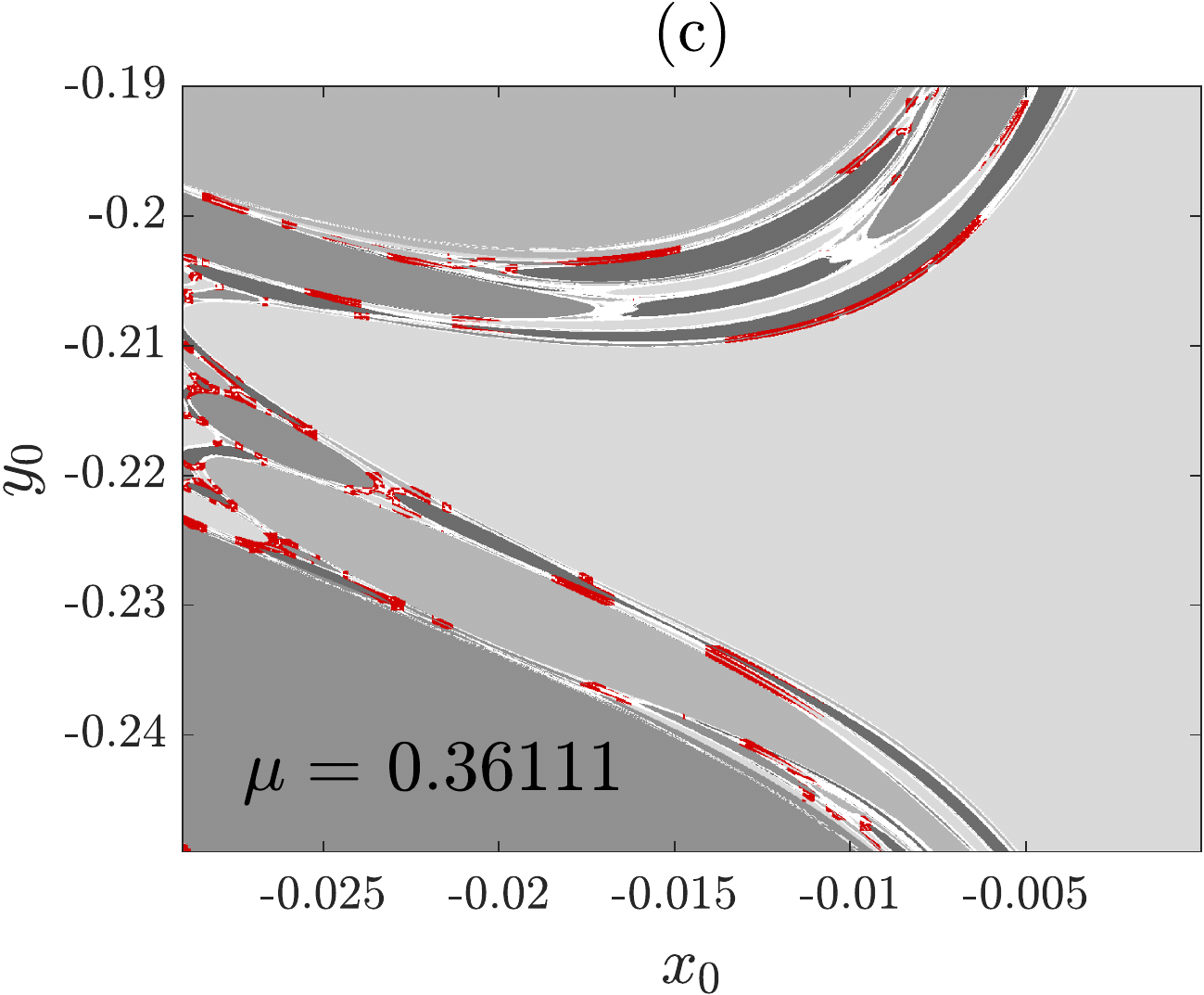}
	\quad
	\includegraphics[width=0.48\textwidth]{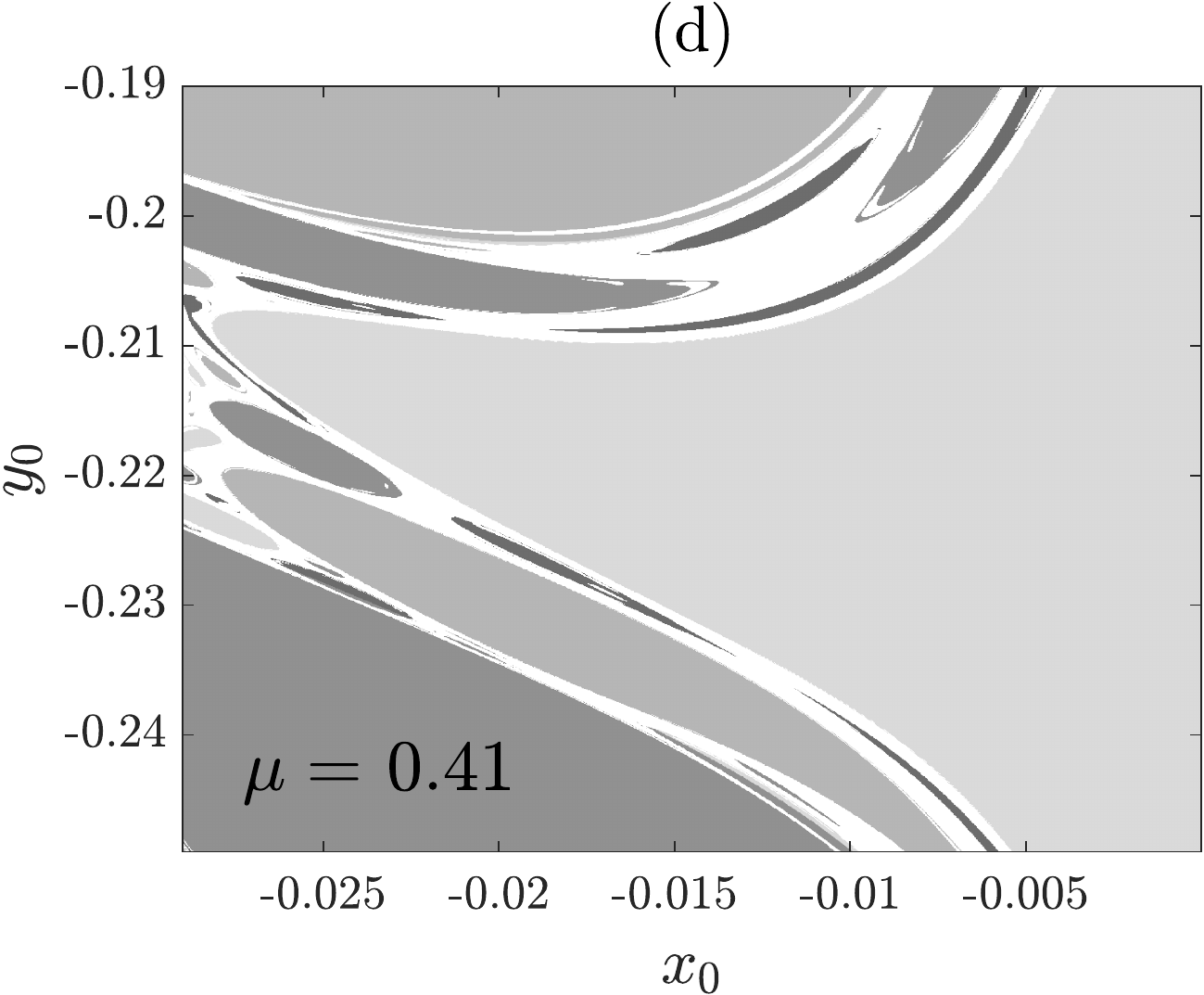}
	
	\caption{Exit basins (gray scale) as computed for $E_0 = 2.5$, and (a) $\mu = 0.09$, (b) $\mu = 0.24556$, (c) $\mu = 0.36111$ and (d) $\mu = 0.41$. We plot in red the Wada points detected by the \textit{merging method} when the fattening parameter value is $r = 6$. Finally, the white basin represents particles trapped in the attractors.}
	\label{fig:9}
\end{figure*}

The fractions of Wada points $f_W$ and trapped particles $f_t$ are displayed versus the dissipation parameter in Figs.~\ref{fig:8}(a) and \ref{fig:8}(b), respectively. In light of the results, the method converges for $r = 6$. Firstly, note that the value of $f_W$ grows slightly when the dissipation is weak enough that the trapped particle fraction is negligible, e.g., roughly, $f_t < 10^{-3}$. Here the proposed mechanism drives the formation of new detectable Wada points and $f_W$ reaches a maximum value, as indicated by a red dashed line in both figures. However, as the dissipation is increased and strong enough to trap a substantial fraction of particles, the number of Wada points decreases drastically. In particular, as already explained above, a white basin appears and the probability of detecting Wada points decreases, since a colored boundary point is less likely to be surrounded by the other three remaining colors. Finally, the fraction of Wada points tends to vanish, $f_W \to 0$, when the dissipation exceeds $\mu_m = 0.39$, which is considered as strong dissipation for $E_0 = 2.5$. We indicate the beginning of the strong dissipative regime by a black dashed line in both figures. This result is in perfect agreement with the findings obtained in Sec.~\ref{sec:3}, where we prove for $\mu > \mu_m$ that all fractal boundaries are removed and replaced by a new smooth basin representing trapped particles. Hence, for $\mu > \mu_m$ all detected boundary points are non-Wada, so that $f_W = 0$. The whole process of basin transformation described here is illustrated in Fig.~\ref{fig:9}, where the appearance of Wada points under weak dissipation and their drastic destruction in the strong dissipation regime can be appreciated.

\section{Conclusions} \label{sec:5}

Along this paper, we have studied the dissipation effects on a three-dimensional open Hamiltonian system. Specifically, in the conservative regime, the basin boundaries undergo a topological metamorphosis at the critical energy $E_c \approx 2.25$. Then, the Wada basins become non-Wada basins when this critical energy is surpassed. In this context, we argue that non-Wada basins expected in the conservative case (for $E>E_c$) can transform themselves into partially Wada basins in some situations where dissipation is able to reduce the system energy below $E_c$. As a matter of fact, we confirm the latter and detect how Wada points emerge on the basin boundaries under weak dissipation, whereas strong dissipation prevents the formation of Wada points. This phenomenology has no counterpart in dissipative two-dimensional open Hamiltonian systems, because there exists no critical energies related to a basin topological metamorphosis. As a result of this research, we claim that the Wada property is driven, enhanced and, consequently, structurally stable under weak dissipation in three-dimensional open Hamiltonian systems.

The analysis carried out here may be useful in other three-dimensional models in which the aim is to increase their predictability. Since a control parameter might exceed a certain threshold value, an unpredictability source such as a Wada boundary can be eliminated. However, we have proved that this control parameter can also be modified by the energy loss over time. In this situation, the appearance of Wada basins would again hinder the predictability. For instance, an application where the friction has crucial consequences might be the three-dimensional chaotic open flows of active particles. They can model from the dynamics of pollutants in the atmosphere (chemical activity) to the plankton population dynamics in an ocean (biological activity) \cite{demoura2004}. Moreover, there exists a rising interest in soft Lorentz gases because they are able to simulate electronic transport in graphene-like structures \cite{klages2019}. Diffusion phenomenon in such systems can be modeled by means of three-dimensional open Hamiltonian systems arranged in complex lattices, so a weak dissipation can hinder their predictability. Finally, it is worth mentioning that dissipation is not the only factor that can drive and enhance Wada structures. In fact, some systems have been proven to exhibit the Wada property when time delays are involved \cite{daza2017}. Thus, dissipation is one more factor to be taken into account when predicting the behavior of dynamical systems, specially, in three dimensions.

\appendix
\section{Derivation and computation of the fractal dimension $D_c$ of the chaotic saddle.} \label{sec:6}

Since the chaotic saddle is the intersection of the stable and unstable manifolds \cite{aguirre2009}, whose dimensions can be denoted as $D_s$ and $D_u$, respectively, then the expression \begin{equation} D_c = D_s + D_u - N = 2D_s - 5 \label{eq:A.1} \end{equation} is satisfied, where $D_s = D_u$ due to the symplectic nature of the Hamiltonian under study \eqref{eq:1}, and $N = 5$ is the dimension of the phase space of the conservative system. On the other hand, if $d$ is the dimension of the exit basin boundaries, and taking into account that such boundaries can be described as the intersection between the stable manifold of the chaotic saddle and the proposed initial condition plane, whose dimension is $D = 2$, then we have \begin{equation} d = D + D_s - N = 2 + D_s - 5. \label{eq:A.2} \end{equation} Finally, we relate the dimensions of the chaotic saddle and the basin boundary by operating Eqs.~(\ref{eq:A.1}) and (\ref{eq:A.2}), yielding \begin{equation} D_c = 2d + 1. \label{eq:A.3} \end{equation}

The latter relation enables the dimension $D_c$ to be computed directly if $d$ is known, which is obtained by the uncertainty exponent algorithm as follows \cite{mcdonald1985}. We compute the exit given an arbitrarily initial condition $(x_0, y_0)$ and also given four additional trajectories starting from slightly perturbed initial conditions, such as $(x_0+\varepsilon, y_0)$, $(x_0-\varepsilon, y_0)$, $(x_0, y_0+\varepsilon)$ and $(x_0, y_0-\varepsilon)$, where $\varepsilon \ll 1$. If all five exits coincide, we denote the initial condition as certain. Otherwise, we say the initial condition is uncertain. Repeating this process for many initial conditions and several values of the perturbation, we obtain that the fraction of uncertain initial conditions behaves as $f(\epsilon) \sim \varepsilon^{D-d} = \varepsilon^{2-d}$. Taking decimal logarithms, we obtain $\log_{10} f(\epsilon) = (2-d) \log_{10} \varepsilon + k$, where $k$ is a constant. This formula allows us to compute $d$ by a linear fit of $\log_{10} f(\varepsilon)$ versus $\log_{10} \varepsilon$.

Two extreme situations can be distinguished here. The unpredictability is maximal when $d = D = 2$, since the basin boundary densely occupies the entire space of initial conditions, and even under minimal system perturbations, the fraction of uncertain conditions does not decrease, i.e., $f(\varepsilon) \sim 1$. In this situation, $D_c = N = 5$ holds, implying that the chaotic saddle densely occupies the entire phase space. On the other hand, the unpredictability is minimal when the basin boundary is completely smooth, $d = 1$, and consequently $D_c = 3$. These two extreme cases have not been observed in any of the simulations carried out, so it is expected that the values of the fractal dimensions are $1<d<2$, and hence $3<D_c<5$. This result is in agreement with those displayed in Fig.~\ref{fig:4}.

For the sake of clarity, we only use the basin region $x_0, y_0 \in [-0.1, 0.1]$ to speed up the uncertainty exponent algorithm. Specifically, for each pair of values $E_0$ and $\mu$, we simulate as many initial conditions as necessary to reach one hundred uncertain conditions and, in addition, repeat the process for fifteen values of the perturbation $\varepsilon \in [10^{-4}, 10^{-1}]$.

\end{document}